\documentclass[journal,a4paper]{IEEEtran}
\UseRawInputEncoding

\usepackage{cite}
\usepackage{amsmath,amssymb,amsfonts}

\usepackage{graphicx}
\usepackage{textcomp}
\usepackage{xcolor}

\usepackage{algorithm}
\usepackage{algpseudocode}

\usepackage{float}
\usepackage{caption}
\usepackage{subcaption}
\usepackage{url}
\usepackage{paralist}

\usepackage{tikz}
\usetikzlibrary{arrows.meta,calc,decorations.pathmorphing}
\usepackage[scaled=0.92]{helvet}

\renewcommand{\baselinestretch}{0.96}

\usepackage{soul}

\usepackage[table]{xcolor}
\usepackage{booktabs,tabularx}

\usepackage{ragged2e} 
\newcolumntype{L}{>{\RaggedRight}X} 

\usepackage{comment}
\usepackage{makecell} 
\usepackage{multirow}

\begin{document}

\title{Overload-Robust Latency in 5G-TSN: A HoL-Enhanced Hybrid Lyapunov Approach for 3GPP Indoor Factory Environments}

\author{Kouros Zanbouri,~\IEEEmembership{Graduate Student Member,~IEEE,}
        Md Noor-A-Rahim,~\IEEEmembership{Senior Member,~IEEE,}
        Cormac J. Sreenan,~\IEEEmembership{Senior Member,~IEEE,}
        and Dirk Pesch,~\IEEEmembership{Senior Member,~IEEE}%
\thanks{The authors are with the School of Computer Science and Information Technology, University College Cork, Cork, Ireland (e-mail:
k.zanbouri@cs.ucc.ie).}}

\maketitle

\begin{abstract}
Private 5G networks are a key enabler for flexible industrial automation, especially when used in conjunction with Time-Sensitive Networking (TSN) technology. In this context, radio schedulers must multiplex safety-critical control traffic with bandwidth-hungry sensing streams over a fixed spectrum allocation. This paper proposes a Head-of-Line (HoL) Enhanced Hybrid Lyapunov scheduler for 5G-TSN networks that augments a drift-plus-penalty queue-stability core with an explicit head-of-line delay term and a class-isolation mechanism. The scheduler is evaluated in a 3GPP Indoor Factory scenario with standardized 3GPP fading, spatial consistency, and clutter blockage, using Automated Guided Vehicles (AGVs) generating concurrent URLLC, eMBB, and mMTC flows mapped to dedicated QoS-flow bearers. A fleet-size sweep of 5--30 AGVs on a fixed 20\,MHz carrier reveals a scheduler-independent capacity threshold at approximately 12 vehicles, verified by resource-block saturation. Below the threshold, the proposed scheduler is competitive with the strongest delay-aware baselines and its head-of-line term halves the URLLC deadline-miss ratio relative to the plain Lyapunov formulation. Beyond the threshold, it degrades selectively where the baselines collapse: at $2.5\times$ overload it delivers $1.8\times$ more URLLC traffic than the proportional-fair and delay-budget-aware baselines with a $\approx 4$--$7\times$ shorter 99th-percentile latency, resolving the capacity shortfall in favour of the critical classes instead of spreading it across the traffic mix, at a quantified cost in aggregate cell throughput. The results position Lyapunov-based scheduling as an attractive overload-robustness mechanism for industrial 5G deployments that must remain dependable under unexpected load conditions.
\end{abstract}

%


\section{Introduction}

Modern manufacturing increasingly depends on communication that is not merely fast but also predictable. Under the Industry 4.0 paradigm, production cells are reconfigured frequently, control loops are distributed across the plant rather than concentrated in a single controller, and machines, products, and operators exchange state information continuously, while the more recent Industry 5.0 concept retains these objectives and adds human-centricity, sustainability, and resilience to them \cite{Ivanov04032023}. Such arrangements place the network itself on the critical path of production, so that a data frame transmitted late constitutes a failed operation rather than a degraded service. That burden is carried today almost entirely by wired industrial Ethernet. Profinet, EtherCAT, and comparable protocols each achieve the required determinism, but each does so through its own proprietary extensions of the Ethernet stack, so that a plant assembled from the equipment of several vendors ends up operating a corresponding number of mutually incompatible time-critical networks \cite{9065178}. Time-Sensitive Networking (TSN) was standardized to remove this fragmentation, substituting a common family of IEEE standards for clock synchronization, resource reservation, scheduled transmission, and traffic shaping, through which bounded end-to-end latency and negligible congestion loss are obtained on commodity Ethernet hardware \cite{10735349}. While nothing in these standards presupposes Ethernet as the underlying medium, Ethernet remains the medium on which they are almost always deployed.

Within industrial environments, 5G is predominantly deployed as a private 5G network, referred to within 3GPP terminology as a Non-Public Network (NPN), in which the radio access network, the core, and the licensed spectrum are installed on the factory premises and operated by the enterprise itself rather than shared with public subscribers \cite{10667526}. Unlike a public network, however, a private deployment is provisioned with a fixed and comparatively narrow spectrum allocation, typically a single dedicated carrier, that cannot be widened in response to transient demand and must consequently accommodate the entire traffic mix of the factory. In the context of Industry 4.0, 5G and TSN are increasingly viewed as complementary technologies rather than competing solutions. While TSN provides deterministic communication and network convergence for time-critical industrial applications, 5G extends these capabilities to wireless environments through Ultra-Reliable Low-Latency Communications (URLLC), enabling flexible deployment of mobile robots, Automated Guided Vehicles (AGVs), and other wireless industrial devices. Significant standardization efforts, particularly within the 3rd Generation Partnership Project (3GPP), have therefore focused on enabling seamless 5G-TSN integration \cite{10584103}. Throughout this work we consider exclusively the private 5G-TSN setting, in which a single enterprise-operated carrier of fixed bandwidth must serve the entire fleet of industrial devices deployed on the factory floor.

While the potential benefits of 5G-TSN integration are widely accepted, achieving deterministic communication within a practical private deployment remains a challenging task. The underlying difficulty, however, does not stem from traffic heterogeneity in itself, since 5G was designed from the outset to multiplex dissimilar service classes and already provides network slicing, 5QI-based QoS flow differentiation, admission control, and configured grants for precisely this purpose. Rather, the difficulty arises because the guarantee furnished by these mechanisms differs in kind from the guarantee that TSN requires. The QoS framework standardized by 3GPP is inherently statistical, in that each 5QI specifies a Packet Delay Budget to be satisfied with a target reliability rather than enforced as a deterministic worst-case bound, and in that slicing and admission control are provisioned at flow-establishment time on the basis of averaged traffic descriptors. TSN determinism, in contrast, is strictly contractual. Once the 5G system is exposed to the Centralized Network Configuration (CNC) entity as a single logical IEEE 802.1 bridge spanning the device-side and network-side TSN translators \cite{3gpp23501, 10584103}, the residence time advertised by that bridge becomes a hard input to the computation of the gate schedules governing the adjacent wired segments. A residence time that holds only in distribution is therefore not a valid input to that computation, since a single overrun does not merely degrade an individual flow, but invalidates the transmission windows already reserved by the downstream bridges and, consequently, the determinism of the entire end-to-end path. The burden of converting a fading, capacity-constrained air interface into a bridge port that exhibits a bounded residence time thus falls almost entirely upon the gNB Medium Access Control (MAC) scheduler, which remains the single element of this QoS chain deliberately left unspecified by 3GPP, and which co-simulation studies of converged deployments consistently identify as the dominant contributor to end-to-end latency \cite{10333533}.

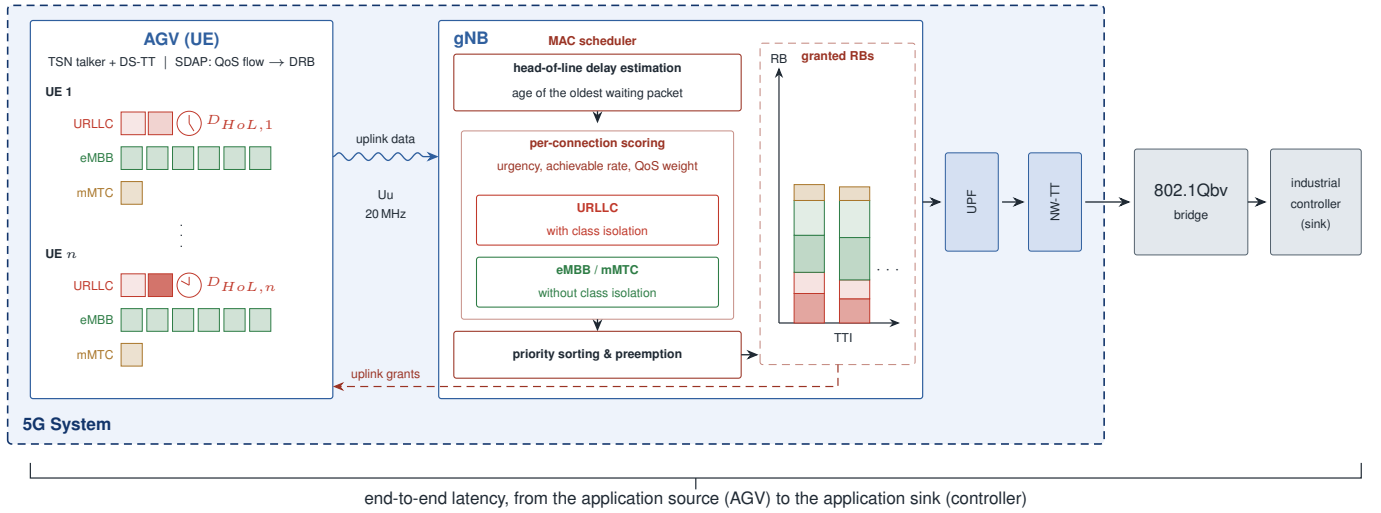
\begin{figure*}[!t]
     \centering
%
%
%
%
%
\definecolor{bInk}   {HTML}{1F2933}%
\definecolor{bWire}  {HTML}{55636F}%
\definecolor{bWireF} {HTML}{E3E8ED}%
\definecolor{b5gs}   {HTML}{12356B}%
\definecolor{b5gsF}  {HTML}{EEF3FB}%
\definecolor{bNode}  {HTML}{2F5F9E}%
\definecolor{bNodeF} {HTML}{D5E0F2}%
\definecolor{bSched} {HTML}{9A3324}%
\definecolor{bSchedF}{HTML}{F7E2DC}%
\definecolor{bCrit}  {HTML}{C0392B}%
\definecolor{bEmbb}  {HTML}{2E7D46}%
\definecolor{bMmtc}  {HTML}{A9762A}%
\definecolor{bCtrl}  {HTML}{6F3BA9}%
\definecolor{bCtrlF} {HTML}{EDE4F7}%
\begin{tikzpicture}[x=1mm,y=1mm,
    font=\sffamily,
    every node/.style={inner sep=0pt,outer sep=0pt},
    blk/.style={draw=bNode,fill=bNodeF,line width=.35,rounded corners=1},
    wblk/.style={draw=bWire,fill=bWireF,line width=.35,rounded corners=1},
    stg/.style={draw=bSched,fill=white,line width=.4,rounded corners=1},
    lbl/.style={font=\sffamily\scriptsize,text=bInk,align=center},
    slbl/.style={font=\sffamily\tiny,text=bInk,align=center},
    flow/.style={draw=bInk,line width=.5,-{Stealth[length=1.6mm,width=1.3mm]}},
    ctrl/.style={draw=bCtrl,line width=.4,densely dashed,
                 -{Stealth[length=1.4mm,width=1.1mm]}}]

\newcommand{\holpkt}[5]{%
  \draw[draw=bCrit,fill=bCrit!12,line width=.3] (#1,#2-1.5) rectangle (#1+3.2,#2+1.5);
  \draw[draw=bCrit,fill=#4,line width=.3] (#1+3.6,#2-1.5) rectangle (#1+6.8,#2+1.5);
  \draw[draw=bCrit,fill=white,line width=.32] (#1+9,#2) circle (1.6);
  \draw[draw=bCrit,line width=.32] (#1+9,#2) -- (#1+9,#2+1.1);
  \draw[draw=bCrit,line width=.32] (#1+9,#2) -- ($(#1+9,#2)+(#3:1.2mm)$);
  \node[slbl,text=bCrit,anchor=west] at (#1+11.2,#2) {#5};}

\draw[draw=b5gs,fill=b5gsF,line width=.6,densely dashed,rounded corners=1.5]
      (1,10) rectangle (146,68);
\node[lbl,text=b5gs,anchor=west,font=\sffamily\scriptsize\bfseries]
      at (3,12.4) {5G System};

\draw[draw=bNode,fill=white,line width=.45,rounded corners=1]
      (4,16) rectangle (44,66);
\node[lbl,font=\sffamily\scriptsize\bfseries,text=bNode] at (24,63.4) {AGV (UE)};
\node[slbl] at (24,60.4) {TSN talker + DS-TT \ \textbar\ \ SDAP: QoS flow $\rightarrow$ DRB};

\node[slbl,anchor=west,font=\sffamily\tiny\bfseries] at (6,56.6) {UE 1};
\node[slbl,anchor=east,text=bCrit] at (15,52.4) {URLLC};
\holpkt{16}{52.4}{-58}{bCrit!22}{$D_{HoL,1}$}
\node[slbl,anchor=east,text=bEmbb] at (15,47.8) {eMBB};
\foreach \i in {0,...,5}{\draw[draw=bEmbb,fill=bEmbb!22,line width=.3]
      (16+\i*3.4,46.3) rectangle (18.8+\i*3.4,49.3);}
\node[slbl,anchor=east,text=bMmtc] at (15,43.2) {mMTC};
\draw[draw=bMmtc,fill=bMmtc!22,line width=.3] (16,41.7) rectangle (18.8,44.7);

\node[slbl] at (24,38.6) {$\vdots$};

\node[slbl,anchor=west,font=\sffamily\tiny\bfseries] at (6,35.2) {UE $n$};
\node[slbl,anchor=east,text=bCrit] at (15,31) {URLLC};
\holpkt{16}{31}{152}{bCrit!70}{$D_{HoL,n}$}
\node[slbl,anchor=east,text=bEmbb] at (15,26.4) {eMBB};
\foreach \i in {0,...,5}{\draw[draw=bEmbb,fill=bEmbb!22,line width=.3]
      (16+\i*3.4,24.9) rectangle (18.8+\i*3.4,27.9);}
\node[slbl,anchor=east,text=bMmtc] at (15,21.8) {mMTC};
\draw[draw=bMmtc,fill=bMmtc!22,line width=.3] (16,20.3) rectangle (18.8,23.3);

\draw[draw=bNode,line width=.5,decorate,
      decoration={snake,amplitude=.5mm,segment length=2.6mm,post length=1.8mm},
      -{Stealth[length=1.6mm,width=1.3mm]}] (44,48) -- (58,48);
\node[slbl,anchor=south] at (51,49.5) {uplink data};
\node[slbl] at (51,43) {Uu};
\node[slbl] at (51,40.4) {20\,MHz};

\draw[draw=bNode,fill=white,line width=.45,rounded corners=1]
      (58,16) rectangle (122,66);
\node[lbl,anchor=west,font=\sffamily\scriptsize\bfseries,text=bNode]
      at (60,63.4) {gNB};
\node[slbl,text=bSched,anchor=west,font=\sffamily\tiny\bfseries]
      at (72.5,63.4) {MAC scheduler};

\node[stg,minimum width=38mm,minimum height=7.5mm] at (79,57.8) {};
\node[slbl,font=\sffamily\tiny\bfseries] at (79,59.6)
      {head-of-line delay estimation};
\node[slbl] at (79,56.3) {age of the oldest waiting packet};
\draw[flow] (79,54) -- (79,52.4);

\draw[draw=bSched!55,line width=.35,rounded corners=1] (61,26.5) rectangle (97,51.4);
\node[slbl,text=bSched,font=\sffamily\tiny\bfseries] at (79,49.7)
      {per-connection scoring};
\node[slbl,text=bSched] at (79,46.6) {urgency, achievable rate, QoS weight};
\node[stg,draw=bCrit,minimum width=32mm,minimum height=6.6mm] at (79,39.6) {};
\node[slbl,text=bCrit,font=\sffamily\tiny\bfseries] at (79,41.2) {URLLC};
\node[slbl,text=bCrit] at (79,38.2) {with class isolation};
\node[stg,draw=bEmbb,minimum width=32mm,minimum height=6.6mm] at (79,31.4) {};
\node[slbl,text=bEmbb,font=\sffamily\tiny\bfseries] at (79,33) {eMBB / mMTC};
\node[slbl,text=bEmbb] at (79,30) {without class isolation};
\draw[flow] (79,26.5) -- (79,24.9);

\node[stg,minimum width=38mm,minimum height=6.2mm] at (79,21.8) {};
\node[slbl,font=\sffamily\tiny\bfseries] at (79,21.8) {priority sorting \& preemption};
\draw[flow] (98,21.8) -- (100.3,21.8);

\draw[draw=bInk,line width=.45,-{Stealth[length=1.4mm,width=1.1mm]}]
      (103,26) -- (103,59.8);
\draw[draw=bInk,line width=.45,-{Stealth[length=1.4mm,width=1.1mm]}]
      (103,26) -- (119,26);
\node[slbl,anchor=south] at (103,60.4) {RB};
\node[slbl,anchor=north] at (111.5,24.6) {TTI};
\foreach \x/\r/\lr/\e/\f/\m in {105/3.9/2.8/4.9/4.6/2.1, 111/3.2/2.5/5.6/4.9/1.8}{%
  \draw[draw=bCrit,fill=bCrit!45,line width=.3] (\x,26) rectangle (\x+4,26+\r);
  \draw[draw=bCrit,fill=bCrit!14,line width=.3] (\x,26+\r) rectangle (\x+4,26+\r+\lr);
  \draw[draw=bEmbb,fill=bEmbb!30,line width=.3] (\x,26+\r+\lr) rectangle (\x+4,26+\r+\lr+\e);
  \draw[draw=bEmbb,fill=bEmbb!14,line width=.3] (\x,26+\r+\lr+\e) rectangle (\x+4,26+\r+\lr+\e+\f);
  \draw[draw=bMmtc,fill=bMmtc!25,line width=.3] (\x,26+\r+\lr+\e+\f) rectangle (\x+4,26+\r+\lr+\e+\f+\m);}
\node[slbl] at (117.6,33) {$\cdots$};
\draw[draw=bSched!60,line width=.35,densely dashed,rounded corners=1]
      (100.5,20.5) rectangle (121,63);
\node[slbl,text=bSched,font=\sffamily\tiny\bfseries] at (110.8,61.2) {granted RBs};

\draw[draw=bSched,line width=.45,densely dashed,
      -{Stealth[length=1.5mm,width=1.2mm]}]
      (110.8,20.5) -- (110.8,17.6) -- (44,17.6);
\node[slbl,anchor=south,text=bSched] at (51,18.4) {uplink grants};

\draw[flow] (122,42) -- (124.5,42);
\node[blk,minimum width=7mm,minimum height=13mm] at (128.5,42) {};
\node[slbl,rotate=90] at (128.5,42) {UPF};
\draw[flow] (132.5,42) -- (135.5,42);
\node[blk,minimum width=7mm,minimum height=13mm] at (139.5,42) {};
\node[slbl,rotate=90] at (139.5,42) {NW-TT};
\draw[flow] (143.5,42) -- (149.5,42);

\node[wblk,minimum width=15mm,minimum height=14mm] at (157.5,42) {};
\node[lbl] at (157.5,43.6) {802.1Qbv};
\node[slbl] at (157.5,40.2) {bridge};
\draw[flow] (165,42) -- (167.5,42);
\node[wblk,minimum width=12mm,minimum height=14mm] at (174,42) {};
\node[slbl] at (174,44.6) {industrial};
\node[slbl] at (174,41.8) {controller};
\node[slbl] at (174,39) {(sink)};

\draw[draw=bInk,line width=.45] (4,7.5) -- (4,5.5) -- (180,5.5) -- (180,7.5);
\draw[draw=bInk,line width=.45] (92,5.5) -- (92,4);
\node[lbl,anchor=north] at (92,3.7)
      {end-to-end latency, from the application source (AGV) to the application sink (controller)};

\end{tikzpicture}
     \caption{Converged 5G-TSN uplink path. The 5G System is bridged into the wired TSN segment through the device-side and network-side translators. Each AGV generates URLLC, eMBB, and mMTC flows carried on dedicated DRBs, the clock on each head-of-line URLLC packet marking that connection's waiting time $D_{HoL,c}(t)$. Backlogs accumulate in the UE, while the gNB MAC scheduler one hop away scores them and returns the per-TTI resource grant.}
     \label{fig:tsn-scheduler}
\end{figure*}

Fig.~\ref{fig:tsn-scheduler} locates this burden within the converged uplink path and shows why it is not discharged by the existing QoS machinery. Two characteristics of the industrial setting are critical, and both act through the scheduler. First, the conflicting traffic classes are co-resident on the same device, as a single AGV concurrently generates both a monitoring video stream and a closed-loop control channel. Slicing, admission control, and spatial reuse, which inherently operate between slices and devices, therefore cannot separate these flows, and the required isolation must instead be enforced among the logical channels sharing the grant issued to that individual UE. Second, the packets accumulate in the per-DRB queues of the UE, whereas the resource that drains them is granted one hop away by the gNB once per Transmission Time Interval (TTI), so that the scheduler governs a backlog it does not itself hold and must act on through the grant alone. Compounding both characteristics, the radio link differs from the wired segments of the TSN path in the nature of its capacity. A wired Ethernet port transmits at a constant rate with negligible loss, so the gate schedules of the adjacent bridges are computed against a
known rate, and bounded latency follows from reserving and shaping the offered traffic. On the radio link the rate itself varies, since path loss, Rician fast fading, and transient clutter blockage alter the achievable rate of each device on a per-TTI basis.

Over-provisioning can mask this variation, keeping the rate that remains in a deep fade sufficient for time-aware shaping, but on a fixed private carrier the margin is paid for in the number of devices served. Once the fleet exceeds the size that margin permits, whether a critical packet meets its bound is decided by the scheduler's per-TTI allocation. Under these conditions the volume asymmetry between the traffic classes becomes decisive. While eMBB applications, such as continuous high-definition video streaming for factory monitoring, demand sustained high-throughput capacity and generate massive queue backlogs, URLLC applications, such as closed-loop robotic controls in TSN, rely on strict, deterministic latency bounds. In traditional queue-aware scheduling paradigms, the sheer volume of eMBB payloads can overwhelm the fixed physical resource block allocation of the private carrier, leading to severe bufferbloat. Separating the classes into their own queues does not by itself prevent this. Each class of an AGV is held in its own DRB queue, as each traffic class of a TSN bridge is held in its own transmission queue, but all of these queues drain through resource blocks that a single gNB scheduler allocates across every connection in the cell. A scheduler that ranks connections by backlog therefore awards those blocks to the large and persistent eMBB queues, and a URLLC packet that shares no buffer with eMBB traffic still waits for want of a grant. The time-aware shaping that TSN applies at bridge egress does not remove this delay, because in a 5G system bridge it is realized by hold-and-forward buffering at the DS-TT and NW-TT ports \cite{10584103}, which can hold back a frame that arrives before its gate opens but cannot release one that the radio has not yet delivered. When the grant arrives too late, the tail latency of the critical class exceeds the residence time advertised for the bridge and the delay budget on which the adjacent gate schedules depend.

Two questions therefore determine the design of the scheduler. The first concerns what the scheduling metric must measure. Legacy policies rank connections by channel quality or by throughput history \cite{9773317}, so that Maximum Carrier-to-Interference scheduling maximizes cell capacity while starving edge users, and Proportional Fair balances instantaneous rate against a historical average that says nothing about how long any individual packet has already waited. QoS-aware refinements of these heuristics scale the same underlying metric by class weights and delay urgency \cite{11309032}, by rate demands \cite{9858321}, or by retransmission failure rates \cite{10679993}, yet because they still allocate resources through localized throughput scaling rather than temporal isolation, they cannot bound the worst-case delay of a critical packet, and under congestion they degrade into precisely the bufferbloat that the converged setting cannot tolerate. What the industrial case requires is not a fairer division of throughput but a bound on the tail, which we formalize in Section~\ref{sec:hol-enhanced-lyapunov} as the minimization of the maximum waiting time incurred by any critical connection subject to its delay budget.

The second question concerns the framework within which such a bound can be pursued, and it is here that Lyapunov optimization becomes attractive. Both the arrivals and the service rate in this scenario are stochastic, the former because the traffic mix is bursty and the latter because Rician fading and clutter blockage alter the achievable rate on a per-TTI basis, and the drift-plus-penalty framework is formulated for exactly this situation, since it establishes queue stability under stochastic arrivals and stochastic service without requiring the arrival statistics to be known in advance or the channel to be predicted \cite{neely2010stochastic}. Minimizing the drift bound moreover reduces the allocation problem to a max-weight comparison evaluated independently per connection at each TTI, so that the resulting control law scales linearly with the number of active connections and executes comfortably within the 0.5\,ms TTI budget, whereas prediction-based or learning-based alternatives introduce training and inference costs that a private industrial deployment must absorb at the network edge. The penalty term further provides a principled place at which system utility can be traded against stability rather than an additional heuristic weight. Crucially for the present work, the framework also localizes the very weakness that must be repaired. Because the classical policy scores each connection by the product of its backlog and its achievable rate, the metric is anchored to queue volume, so the sustained backlog of an eMBB stream arithmetically overwhelms the typical few hundred bytes of a waiting control packet and the scheduler exhibits the threshold blindness described in Section~\ref{sec:hol-enhanced-lyapunov}. The defect is therefore confined to a single term of an otherwise sound formulation, which makes it possible to substitute a delay-based quantity for that term while retaining the stability guarantee of the surrounding framework, rather than discarding the framework in favour of a further heuristic.

To address this, we propose a novel Head-of-Line (HoL) Enhanced Hybrid Lyapunov scheduling framework and evaluate it within a high-fidelity industrial physical-layer model. Our contributions are threefold:
\begin{enumerate} \item \textbf{HoL-Enhanced Hybrid Lyapunov Scheduling:} We develop a scheduling policy that replaces the queue-backlog weight of the drift-plus-penalty framework with a bounded exponential function of the age of each connection's head-of-line packet, and we embed this function in the class-isolated hybrid scheduler of our earlier work \cite{11555379}. Critical traffic retains strict precedence over elastic traffic, while competing critical connections are prioritized at every TTI according to how close their oldest packets are to their delay budgets. This removes the queue-size dominance through which bulk eMBB backlogs starve small, time-critical control packets. Because the policy reacts to the measured age of queued packets and evaluates each connection independently, it requires neither prior knowledge of traffic periodicity nor reserved radio resources.

\item \textbf{Industrial 5G-TSN Evaluation Framework:} We develop a system-level evaluation framework for converged industrial 5G-TSN networks by extending an established 5G network simulator with the 3GPP TR~38.901 Indoor Factory (InF) propagation scenario \cite{3gpp38901}, a Rician line-of-sight component whose $K$-factor evolves consistently with the trajectory of each AGV, and a stochastic model of transient blockage. The framework represents a fleet of AGVs whose URLLC control, eMBB video, and mMTC telemetry flows are mapped to dedicated QoS-flow bearers and delivered over the 5G uplink to an industrial controller behind a TSN bridge, so that scheduling policies can be compared end to end under mobility, fading, and blockage.

\item \textbf{Scalability and Overload Analysis:} We conduct a systematic fleet-size study that compares the proposed policy with traditional and state-of-the-art schedulers from lightly loaded operation to sustained overload. The study identifies two distinct operating regions separated by a capacity threshold that is common to all schedulers: below the threshold, all policies deliver the offered traffic and differ mainly in deadline compliance, whereas beyond it the scheduling policy determines whether critical traffic continues to be delivered, and the proposed scheduler preserves the delivery of the critical class. The analysis further reveals that mean latency computed over delivered packets misranks schedulers under overload, and that a URLLC latency floor persists even at light load that dynamic uplink scheduling alone does not remove, with direct implications for the dimensioning and evaluation of industrial 5G deployments. 
\end{enumerate}

The rest of this paper is structured as follows. Section \ref{sec:related_work} reviews the existing literature and related research efforts. Section \ref{sec:system_model} introduces the proposed system architecture and provides an overview of the underlying model. The proposed HoL-Enhanced Lyapunov scheduling framework is detailed in Section \ref{sec:hol-enhanced-lyapunov}. Section \ref{sec:Simulation&Methodology} outlines the simulation environment, experimental configuration, and channel emulation procedures. Subsequently, Section \ref{sec:Results} presents the performance assessment and discusses the obtained numerical results. Finally, Section \ref{sec:conclusion&futureWork} summarizes the main contributions of this work and highlights promising directions for future research.

\section{Related Work} \label{sec:related_work}
Prior work relevant to this paper falls into three groups. The first concerns general-purpose 5G MAC scheduling, which supplies the baseline policies but is not concerned with deterministic networking. The second concerns QoS-aware refinements of those policies, and the third, which is the group this work belongs to, concerns scheduling for converged 5G-TSN deployments and the Lyapunov-based methods that offer provable queue stability. We treat each in turn, giving the greatest attention to the last two.

\subsection{General-Purpose 5G MAC Scheduling}
The 5G MAC layer, specifically the scheduler at the gNodeB (gNB), handles the allocation of physical resources to logical channels. At each TTI, the scheduler assigns radio resources to UEs based on scenario-specific Key Performance Indicators (KPIs). Radio resource management ranks UEs based on priority metrics and establishes a mapping between Resource Blocks (RBs) and UEs, taking into account their Quality of Service (QoS) requirements. Scheduling approaches are generally classified into two types. Channel-Independent Scheduling (CIS) applies a static allocation strategy that does not consider the type of traffic or current channel conditions. In contrast, Channel-Dependent Scheduling dynamically allocates resources based on real-time conditions reported by the UEs. Dynamic scheduling relies on feedback such as Channel State Information (CSI), Buffer Status Reports (BSR), and traffic-specific QoS attributes. Using this data, the scheduler computes a metric \( w_{i,j} \) per UE-RB pair. The UE with the highest \( w_{i,j} \) for RB \( j \) gets access during that TTI. The way the metrics are calculated differs depending on the scheduling strategy \cite{9773317}.

Within these paradigms, several traditional scheduling algorithms have been widely adopted in standard 5G deployments, including Round Robin (RR), Maximum Carrier-to-Interference (MaxCI), and Proportional Fair (PF). Round Robin assigns equal portions of transmission time to each user in a circular order and attains strong fairness when users present similar channel conditions and similar packet sizes \cite{8674018}, but it disregards channel quality and queue depth entirely, and comparative evaluations across mixed traffic models report it as the weakest of the standard policies in both throughput and delay \cite{10620389}. Conversely, MaxCI, which is often also called Best-CQI, strictly maximizes system capacity by favoring UEs with the most robust channel conditions, inherently causing resource starvation for edge-cell users and failing to account for traffic deadlines \cite{9773317}. Proportional Fair attempts to strike a balance by scaling instantaneous channel conditions against historical average data rates \cite{9773317}. However, while effective for homogeneous data streams, these legacy schedulers exhibit critical vulnerabilities when confronted with heterogeneous traffic profiles. When massive eMBB payloads are multiplexed with small, time-critical URLLC packets and dense, massive Machine-Type Communication (mMTC) bursts, algorithms like PF and MaxCI inherently favor the volumetric efficiency of eMBB or the channel superiority of specific nodes. Because they lack explicit latency-awareness and deterministic bounding, these standard approaches leave the delay contributed by the radio segment unbounded. In a converged deployment, this is a decisive shortcoming, as co-simulation studies attribute the majority of the end-to-end delay to the 5G segment, since an unbounded delay in 5G misaligns the gate opening and closing times of the TSN switch on the device side \cite{10333533}.

\subsection{QoS-Aware Scheduling in 5G}
A first line of response retains the proportional-fairness or throughput-maximizing structure of these baseline approaches and augments the ranking metric with QoS state, such as delay urgency, guaranteed bit rate fulfillment, 5QI priority, or explicit rate demands. These schemes consistently improve average throughput and average user satisfaction relative to the unmodified baselines, but they share a common limitation, in that prioritization is expressed as a weighting of an aggregate utility rather than as temporal isolation. None of the works discussed in this subsection addresses TSN integration.

Closest to the present setting is QoS-PF \cite{11309032}, which we later adopt as a baseline. Implemented in Simu5G for a smart-factory scenario, it ranks flows by a proportional-fairness metric that divides a composite utility, combining delay urgency, GBR fulfillment, and 5QI-derived priority through tunable per-flow weights, by each flow's moving-average throughput, thereby boosting under-served but delay-critical flows while throttling flows that have already received a disproportionate share of resources.

The remaining schemes vary the same idea along different axes. CQAS \cite{10620389} formulates allocation as a mixed binary integer program over latency targets, reliability parameters, and instantaneous channel quality, reporting throughput gains over Round Robin, best-CQI, and QoS-aware scheduling together with satisfaction of the delay constraints of the real-time applications considered. \cite{10679993} applies causal discovery to a live 5G NR network, and, having established a causal path between throughput and latency, multiplies the PF metric by the inverse of each user's hybrid automatic repeat request (HARQ) retransmission failure rate, reporting gains in efficiency and resource utilization. D-PF \cite{9858321} prioritizes millimeter-wave users that combine low rate demands with favourable channels until their targets are met, after which ranking reverts to channel quality alone, improving aggregate capacity and user satisfaction for bandwidth-intensive services. In each case the evaluation is expressed in terms of aggregate throughput, mean delay, and fairness rather than the tail of the delay distribution, so none of these approaches establishes a worst-case bound for an individual packet, which is the guarantee a TSN bridge must provide.

\subsection{Scheduling for Converged 5G-TSN Networks}
A second and more directly related line of work places the 5G system inside a TSN domain, where the radio segment is no longer evaluated on average throughput but on whether it preserves the determinism of the bridged path. \cite{9065178} analyses multi-user scheduling in a TSN-enabled 5G system for industrial applications using a system-level simulator, and reports that the number of deterministic streams supportable under a fixed latency bound falls sharply as the channel degrades, from 50 devices at a channel quality indicator of 12 to 14 devices at an indicator of 5. The authors further show that prioritizing time-sensitive traffic imposes a throughput penalty of roughly 10 to 20 percent on elastic background traffic while raising its latency and jitter, and they conclude that the TSN functions specified for 5G are not by themselves sufficient to deliver the performance TSN requires, so that the wired and wireless segments cannot be scheduled independently of one another. \cite{10584103} surveys this landscape exhaustively, covering both the standards and the academic literature through which 3GPP exposes the 5G system to a TSN control plane, and identifies the trade-offs involved in jointly configuring and scheduling resources across the 5G and TSN systems as an open research challenge. \cite{10333533} proposed 5GTQ, a QoS-aware 5G-TSN co-simulation framework that models the 5G system as a logical IEEE 802.1 bridge through network-side and device-side TSN translators, following 3GPP Release 17. The framework introduces a QoS mapping mechanism that assigns TSN industrial traffic types to standardized 5QI values across the DC-GBR, GBR, and Non-GBR resource types, and schedules the resulting flows in the 5G system using its DQoS scheduler, which prioritizes traffic according to the mapped 5QI priorities. Evaluating two representative scenarios, the authors show that prioritized TSN flows can achieve end-to-end delays within 3 ms and identify the 5G system as the dominant source of overall latency, motivating tighter integration between TSN and 5G scheduling. This body of work establishes the architecture and the QoS mapping through which TSN traffic reaches the radio bearers, but in each case the scheduling decision itself is left to a priority discipline over the mapped 5QI values.

A more recent collection of works moves beyond the integration architecture and addresses the radio scheduling decision itself. \cite{9940254} considers a hybrid network in which a controller sends periodic control packets to industrial equipment through a TSN bridge and a 5G logical bridge, and in which eMBB streams share the radio resources with the time-sensitive streams. Because the streams are periodic, the instant at which each control packet arrives at the base station can be determined in advance, and every stream is accordingly assigned a window of mini-slots within which its transmission must complete. Resource blocks are then allocated in two stages, first to the time-sensitive streams so as to minimize their latency and afterwards to the eMBB streams so as to maximize throughput, the time-sensitive priority being formed as the product of the fraction of the delay budget already elapsed and the relative quality of the subchannel across the band. The reported worst-case end-to-end delay remains at 0.6\,ms for as many as 70 time-sensitive streams, and more than 100 streams are supported under a 1\,ms constraint. Of the works surveyed here this is the closest in structure to the metric developed in Section~\ref{sec:hol-enhanced-lyapunov}, since it too ranks a stream by a delay term scaled by a channel term. The delay term is nevertheless linear in the fraction of the budget consumed and is evaluated against a deadline known in advance from the period of the stream rather than against the measured age of the packet actually waiting at the head of the queue, and the separation between the classes is expressed as a fixed two-stage ordering rather than as a term within a single score.

\cite{network2030027} asks instead how much of the traffic forwarding required by the TSN control plane can be performed by native radio access network mechanisms rather than by translator functions at the edge of the 5G system, and concludes that a time-invariant logical channel prioritization is not appropriate for TSN streams over the air interface. Every time-sensitive communication QoS flow is required to be mapped to its own radio bearer, so that distinct streams present themselves to the MAC scheduler as separately servable logical channels, and the logical channel priorities are then updated at every TTI in accordance with the current state of the gate control list, a blocking priority being introduced under which a logical channel is not served at all for the corresponding interval. The gating behaviour of the time-aware shaper is thereby reproduced within the scheduler itself. A downlink simulation of three time-sensitive bearers and one best-effort bearer on a single UE shows that the shaping requirements of the gate control list can be met, subject to the condition the authors state explicitly, namely that the radio resources available within one gate control list period suffice to carry the traffic of the corresponding class.

\cite{10110348} couples the mapping and the scheduling problems. A dynamic QoS mapping algorithm clusters flows according to their QoS attributes and assigns the resulting aggregates to 5QI values, after which an adaptive semi-persistent scheduling mechanism reserves resource blocks persistently for the deterministic periodic flows, whose periodicity is supplied by the traffic assistance information, so that those flows transmit without the scheduling request and grant exchange, while the remaining resource blocks are distributed among the other flows by a max-min fair share weighted by the inverse of the priority. Under loads of up to several hundred flows the average end-to-end delay of the time-sensitive traffic is held below 2\,ms and that of the non-time-sensitive traffic below 8\,ms, at a resource utilization above that of static reservation. The authors also report the cost of the approach, in that the proportion of resource blocks which must be reserved grows with the proportion of time-sensitive traffic, and reserving more than the traffic requires wastes capacity.

Taken together, these works supply the architecture, the QoS mapping, and in the three most recent cases a radio scheduling mechanism by which TSN traffic is given preferential treatment on the air interface. 
\cite{network2030027} states the condition explicitly and \cite{10110348} quantifies its cost as the reserved fraction grows. None of these schemes arbitrates between competing critical connections by the measured queueing age of the packet at the head of each queue, and none is evaluated in the regime in which the offered load exceeds the capacity of the cell, which is precisely the regime in which the reservation ceases to be adequate and in which the delay of an individual packet is settled by the order in which the scheduler serves the queues. That regime, and the behaviour of a scheduler within it, is the subject of the present work.

\subsection{Lyapunov-Based Scheduling}
To overcome the fundamental limitations of these heuristic and average-rate-based frameworks, a third line of work turns to rigorous mathematical models capable of enforcing queue stability under stochastic arrivals. Such approaches build on Neely's drift-plus-penalty framework \cite{neely2010stochastic}, which stabilizes network queues while opportunistically optimizing a system utility, and which requires neither prior knowledge of the arrival statistics nor prediction of the channel. The framework has since been carried into 5G resource allocation in several settings. \cite{7430261} applies the drift-plus-penalty theorem to 5G mobile nodes with vertical multi-homing, aggregating several radio access technologies at the network protocol layer in order to maximize aggregate throughput under average power constraints while holding the queues stable. \cite{cmu2.12264} formulates uplink resource allocation for underlaid device-to-device communication as a mixed-integer non-linear problem, and couples Lyapunov optimization with a support vector machine to maximize system capacity subject to bit-error-rate and transmit-power constraints. Both confirm the practical value of the drift-plus-penalty structure for 5G resource allocation, since in neither case must the arrival statistics be known in advance. In both, however, the penalty term carries an average-rate objective, namely aggregate throughput or system capacity, and the queues are accordingly stabilized in the mean rather than bounded packet by packet. Neither therefore speaks to the deadline of the individual head-of-line packet, which is the critical quantity in our converged industrial setting.

Building on this prior work, we previously introduced a Hybrid Lyapunov scheduler \cite{11555379} that couples the drift-plus-penalty backlog term with a strict URLLC priority-isolation booster, demonstrating that queue-stability optimization combined with class isolation can protect critical traffic under load. Nonetheless, existing Lyapunov-based schedulers remain fundamentally anchored to the queue-backlog term, which leaves them vulnerable to volume dominance when bandwidth-intensive elastic traffic is multiplexed with sparse, deadline-critical flows on shared radio bearers. Addressing this limitation while preserving the stability guarantees of the drift-plus-penalty structure motivates the scheduling framework developed in this work, which we detail in the following sections.

\section{System Model} \label{sec:system_model}

\begin{figure*}[!t]
    \centering
    \includegraphics[width=\textwidth]{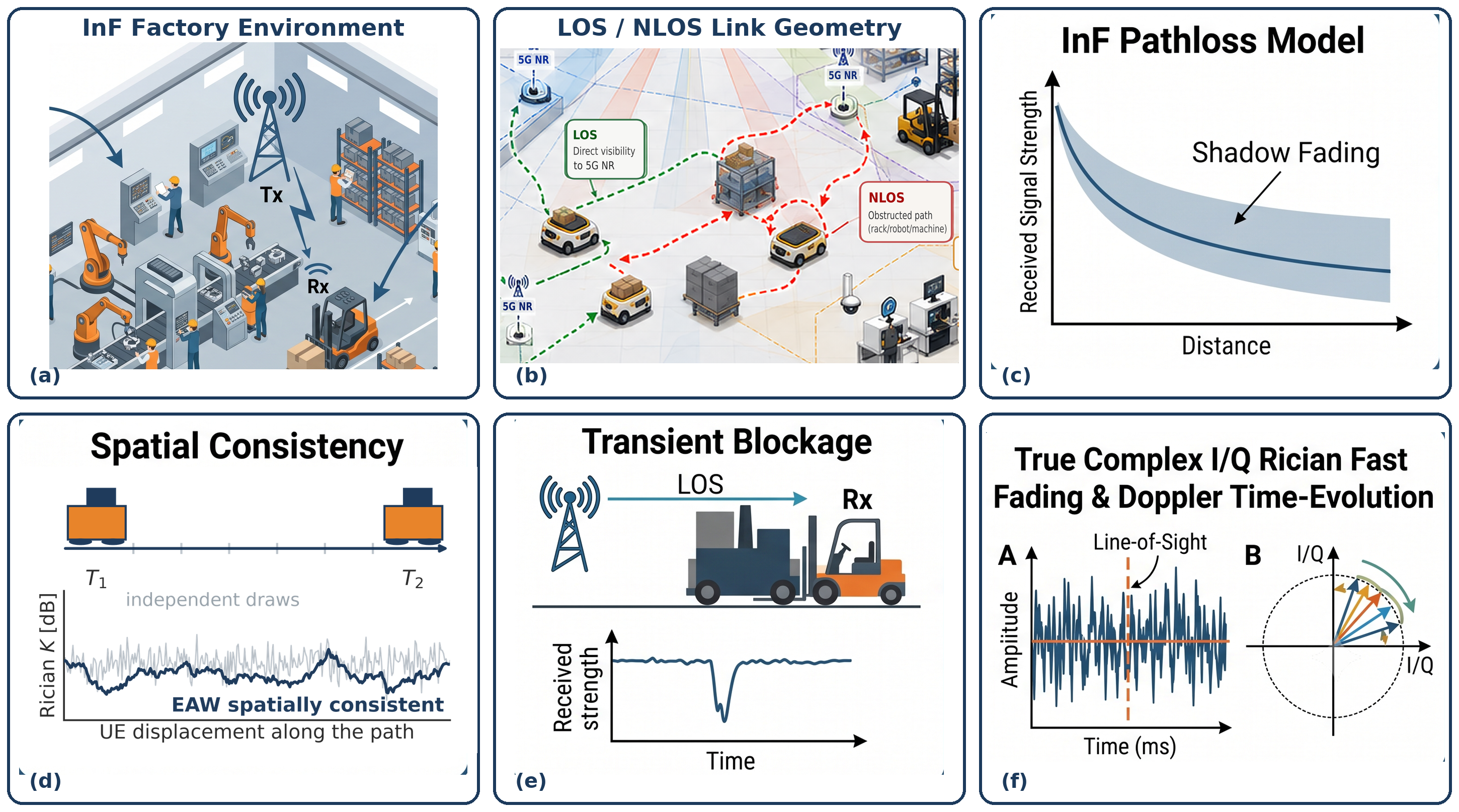}
    \caption{The 3GPP Indoor Factory (InF) scenario and the stochastic propagation mechanisms modeled on it. (a) the dense-clutter factory floor served by a single gNB; (b) the link geometry of the AGV fleet, with line-of-sight (green) and clutter-obstructed non-line-of-sight (red) paths to the 5G NR radio units; (c) the InF path-loss model with log-normal shadow fading; (d) EAW spatial consistency: as an AGV traverses the floor from $T_1$ to $T_2$, the Rician $K$-factor drawn independently at every evaluation point (grey) is replaced by the spatially correlated realization produced by the first-order Gauss-Markov recursion of (\ref{eq:spatially-consistent}) (navy), correlated across the evaluation points marked along the path; (e) a transient clutter blockage event and the resulting drop in received strength; (f) complex I/Q Rician fast fading with Doppler time-evolution, showing the amplitude process (A) and the rotation of the I/Q components (B).}
    \label{fig:inf-environment}
\end{figure*}

This section establishes the mathematical and physical foundations of the converged 5G-TSN network. We first define the overarching multi-tier network architecture. Subsequently, we detail a high-fidelity physical layer model designed to capture the severe stochasticity of the Indoor Factory (InF) environment, incorporating the 3GPP TR~38.901 InF path-loss model, spatial consistency, continuous time-evolving Rician fading and Stochastic Transient Blockage. Fig.~\ref{fig:inf-environment} illustrates the model elements and structures the remainder of this section. 
Taken together, the panels describe a channel whose achievable rate varies on the time scale of individual transmission opportunities, which is the property that the scheduler of Section~\ref{sec:hol-enhanced-lyapunov} must absorb. Finally, we formalize the heterogeneous traffic arrival processes and the corresponding MAC-layer queueing dynamics.

\subsection{Network Architecture} We consider a converged 5G-TSN industrial network architecture operating in Time Division Duplex (TDD) mode. The network consists of a single next-generation NodeB (gNB) serving a set of user equipments (UEs), denoted $\mathcal{U}$, each representing an AGV. Every UE concurrently generates three traffic flows (URLLC, eMBB, and mMTC), which the SDAP layer maps onto three dedicated Data Radio Bearers (DRBs), so that the classes reach the MAC scheduler as separately servable logical channels rather than as a single aggregated bearer \cite{network2030027}. The scheduler therefore operates on the set of logical connections $\mathcal{C}$, where each UE $u \in \mathcal{U}$ contributes one connection per class, so that $\mathcal{C} = \mathcal{C}_{URLLC} \cup \mathcal{C}_{eMBB} \cup \mathcal{C}_{mMTC}$ partitions the \emph{connections} (not the UEs) by traffic class, and we write $u(c) \in \mathcal{U}$ for the UE to which connection $c$ belongs. Time is discretized into uniform Transmission Time Intervals (TTIs) of duration $\Delta t$, indexed by $t \in \{0, 1, 2, \dots\}$. At each TTI $t$, the gNB schedules a finite pool of frequency-domain physical RBs, denoted $B_{total}$, across the active logical connections in $\mathcal{C}$.

\subsection{Industrial Channel and Physical Layer Model} To accurately capture the harsh, highly reflective, and dynamic nature of an industrial factory floor, we extend the realistic channel model of Simu5G \cite{9211504} into a high-fidelity 3GPP-based Indoor Factory (InF) stochastic propagation framework. The framework combines model elements specified in 3GPP TR~38.901 \cite{3gpp38901}, elements already provided by Simu5G, and extensions developed in this work, and the origin of each element is identified in the following.

\subsubsection{InF Path Loss and Spatial Consistency} 
The large-scale propagation is governed by the 3GPP TR 38.901 Indoor Factory (InF) framework \cite{3gpp38901}. Specifically, the base path loss is evaluated via a dedicated function, which we added to the Simu5G implementation of TR~38.901 because the Simu5G release does not include the InF scenario, to apply the equations in Table~7.4.1-1 for varied InF sub-scenarios, such as InF-SL (sparse clutter, low base station), InF-DL (dense clutter, low base station), InF-SH (sparse clutter, high base station), InF-DH (dense clutter, high base station), and InF-HH (high transmitter, high receiver) (Table~7.2-4), utilizing the distance and frequency-dependent coefficients of that table. Concurrently, the function provides the LOS probability, $P_{LOS}(d_{2D})$, of TR 38.901 Table~7.4.2-1, which decays exponentially with the horizontal distance $d_{2D}$ at a rate set by the typical clutter size ($d_{clutter}$) and the clutter density ($r$) and, for the sub-scenarios with a high base station, also by the effective clutter height ($h_{clutter}$), reflecting the physical dimensions and density of the metallic obstacles separating the gNB and the UE. In the evaluation reported in this paper, every gNB-UE link is assigned the LOS state, so that the LOS path-loss expression of Table~7.4.1-1, which is common to all InF sub-scenarios, applies throughout, and obstruction of the LOS path is represented by the transient blockage process of Section~\ref{sec:blockage}.

In highly mobile industrial topologies, such as AGVs navigating the factory floor, drawing channel parameters independently at each discrete time step would cause physically impossible, instantaneous discontinuities in the simulated signal-to-noise ratio (SNR), rendering scheduler evaluations inaccurate. To ensure realistic channel continuity, we apply the EAW spatial consistency procedure, which Simu5G uses for the shadow fading, also to the Rician K-factor.

Let $\Delta d_u(t)$ denote the 3D spatial displacement of UE $u$ since the last update of its K-factor. TR 38.901 describes the spatial correlation of shadow fading by an exponential autocorrelation function of the displacement, and gives the correlation distances for shadow fading and the other large-scale channel parameters, including the K-factor, in Table~7.5-6 (Clause~7.4.4) \cite{3gpp38901}. This correlation is characterized by a scenario-specific decorrelation distance, $d_{cor}$. The shadow-fading procedure of Simu5G evaluates this exponential with an additional coefficient of 0.5 in its argument, and we adopt the same coefficient for the K-factor. Thus, the spatial correlation coefficient $a_u(t)$ utilized in our system is calculated as:

\begin{equation}
    a_u(t) = \exp \left( -0.5 \cdot \frac{\Delta d_u(t)}{d_{cor}} \right)
    \label{eq:EAW-formula}
\end{equation}

To realize this prescribed spatial correlation in a computationally efficient manner, a first-order Gauss-Markov (AR-1) process is employed. Whenever $\Delta d_u(t)$ exceeds $d_{cor}$, the spatially consistent, updated deviation component is computed recursively as:

\begin{equation}
    \mathcal{X}_{new,u}(t) = a_u(t) \cdot \mathcal{X}_{old,u}(t)+ \sqrt{1 - a_u(t)^2} \cdot \mathcal{X}_{drawn,u}(t)
    \label{eq:spatially-consistent}
\end{equation}

In eq. (\ref{eq:spatially-consistent}), $\mathcal{X}_{old,u}(t)$ denotes the previously stored deviation state associated with UE $u$, while $\mathcal{X}_{drawn,u}(t)$ represents a newly generated, statistically independent Gaussian random variable drawn from the original K-factor deviation distribution. The coefficient $a_u(t)$ determines the degree of spatial memory retained from the previous channel state, whereas the normalization term $\sqrt{1-a_u^2(t)}$ preserves the variance and stationarity of the stochastic process. Consequently, the correlation between two consecutive updates is exactly equal to the prescribed spatial correlation coefficient $a_u(t)$, and between updates the stored value is retained.

The updated deviation component is subsequently incorporated into the instantaneous Rician K-factor employed during the SINR computation. To ensure physical validity and prevent negative power ratios, the stochastic deviation is applied in the logarithmic (decibel) domain. Let $K_{dB,u}(t)$ represent the K-factor in decibels for node $u$. The updated state is computed as:
\begin{equation}
K_{dB,u}(t) = \mu_K + \mathcal{X}_{new,u}(t)
    \label{eq:K-factor in decibels for node}
\end{equation}

where $\mu_K$ denotes the scenario-dependent mean K-factor [dB] determined by TR 38.901 Table 7.5-6, and $\mathcal{X}_{new,u}(t)$ introduces the spatially correlated stochastic variation. The instantaneous linear K-factor applied to the channel coefficients is then obtained via standard decibel-to-linear conversion:

\begin{equation}
K_u(t) = 10^{\frac{K_{dB,u}(t)}{10}}
    \label{eq:instantaneous-kfactor}
\end{equation}
This logarithmic formulation strictly guarantees $K_u(t) > 0$ for all $t$, maintaining continuous physical consistency. By employing the updated K-factor in the SINR calculation, the gNB scheduler observes spatially correlated channel conditions rather than unrealistic stochastic oscillations caused by independently regenerated channel parameters.

Furthermore, Simu5G applies the same first-order Gauss-Markov recursion to the large-scale shadow fading component, so that both the K-factor and the shadow fading are spatially consistent. This unified treatment enables the generated channel realizations to exhibit temporally and spatially consistent signal-to-noise ratio (SNR) evolution, following the exponential form of the spatial correlation that 3GPP TR 38.901 specifies for large-scale parameters.

Finally, when an AGV encounters severe physical blockage or transitions from a LOS to a Non-Line-of-Sight (NLOS) propagation condition, the previously stored K-factor state is discarded and a new statistically independent realization is generated. This reset mechanism models the abrupt loss of channel correlation that occurs following significant changes in the propagation environment and prevents unrealistic persistence of LOS channel characteristics across physical obstructions. Consequently, upon re-emerging from the blockage region, the AGV establishes a new independent K-factor state that reflects the altered radio environment.

\subsubsection{Rician Fast Fading and Doppler Time-Evolution}
Due to the dense presence of metallic machinery and reflective surfaces in InF environments, small-scale fading cannot be accurately captured by simple Rayleigh distributions or discrete stochastic jumps. The channel exhibits a dominant, yet dynamically shifting, LOS component mixed with intense multipath scattering, for which TR 38.901 specifies a Rician K-factor under InF LOS conditions (Table~7.5-6) \cite{3gpp38901}. To capture this physical reality, we add a LOS component to the complex In-phase and Quadrature (I/Q) sum-of-sinusoids fading of Simu5G, whose scattered components already evolve continuously with the Doppler frequency. The resulting Rician fading is governed by the instantaneous Rician K-factor $K$, modelled as a log-normal random variable with scenario-dependent mean $\mu_K$ and standard deviation $\sigma_K$ following TR 38.901 Table 7.5-6, and evolved across space via the EAW spatial consistency filter.

Following the classical complex baseband representation of a Rician fading channel, the overall channel coefficient $h_u(t)$ for UE $u$ at simulation time $t$ is modelled as the superposition of a deterministic LOS component and a diffuse NLOS scattering component \cite{10552814}. The weighting factors are determined by the instantaneous Rician K-factor, defined as the ratio between the average powers of the LOS and scattered components \cite{3gpp38901}. Assuming unit average channel power, the LOS and NLOS power contributions become $\frac{K}{K+1}$ and $\frac{1}{K+1}$, respectively. Consequently, the equivalent narrowband channel coefficient is expressed as \cite{10552814}:

\begin{equation}
h_u(t) = \sqrt{\frac{K}{K+1}}\,h_{LOS}(t) + \sqrt{\frac{1}{K+1}}\,h_{NLOS}(t) 
\label{eq:Thechannelcoefficient}
\end{equation}

where $h_{LOS}(t)$ denotes the deterministic LOS component and $h_{NLOS}(t)$ represents the aggregate contribution of the scattered multipath components. Crucially, rather than treating these components as static or independently drawn random variables per TTI, their phase rotations are coupled to the global continuous simulation clock, as in the Simu5G fading model for the scattered component and, in our extension, for the LOS component. In this work, the LOS component is modelled as a unit-magnitude complex phasor with a continuously evolving phase:

\begin{equation}
h_{LOS}(t)=e^{j\phi_{LOS}(t)} \label{eq:h_los}
\end{equation}

while the scattered component is represented by the normalized superposition of $N_p$ independent multipath phasors:
\begin{equation}
h_{NLOS}(t)= \frac{1}{\sqrt{N_p}} \sum_{n=1}^{N_p} e^{j\phi_n(t)}
\label{eq:h_nlos}
\end{equation}
where $N_p$ represents the total number of discrete multipath scatterers, following the classical Jakes Sum-of-Sinusoids (SoS) fading model as implemented in Simu5G.
The phase terms $\phi_{LOS}(t)$ and $\phi_n(t)$ evolve continuously according to the Doppler dynamics described subsequently. This formulation ensures that both the deterministic and scattered components exhibit physically continuous temporal evolution, thereby avoiding the unrealistic channel discontinuities that arise when fading coefficients are independently regenerated at every TTI.

For an AGV moving with velocity $v$ and operating at carrier frequency $f_c$, the maximum Doppler frequency is given by $f_d = \frac{v f_c}{c_0}$, where $c_0$ denotes the speed of light. Consequently, the phase of the specular LOS component evolves continuously according to:

\begin{equation}
    \phi_{LOS}(t) = 2\pi f_d t,
    \label{eq:los_phase}
\end{equation}

which corresponds to the Doppler-induced phase rotation of the LOS path under the assumption that the LOS path arrives along the direction of motion. Similarly, the phase of the $n$-th scattered component is modelled as:

\begin{equation}
    \phi_n(t) = 2\pi \left( f_{d,n} t - f_c \tau_n \right),
    \label{eq:nlos_phase}
\end{equation}

where $f_{d,n} = f_d \cos(\theta_n)$ denotes the Doppler shift associated with the angle of arrival $\theta_n$ of the $n$-th multipath component, drawn uniformly on $[0, \pi]$. $\tau_n$ represents the propagation delay of the $n$-th scattered path, drawn from an exponential distribution with root-mean-square (RMS) delay spread $\tau_{rms}$. The term $2\pi f_c \tau_n$ captures the phase rotation induced by the propagation delay of the corresponding multipath component.

By evaluating the complex coefficient $h_c(t)$ using the continuously evolving phases in eq. (\ref{eq:los_phase}) and eq. (\ref{eq:nlos_phase}), the generated fading process exhibits temporally correlated fading dips and recovery periods characteristic of mobile industrial environments. This continuous evolution prevents unrealistic channel discontinuities caused by independently regenerating fading coefficients at every TTI, thereby enabling the proposed HoL-Enhanced Lyapunov scheduler to be evaluated under physically consistent fading durations and blockage events.

Finally, to emulate the residual Doppler spread introduced by moving machinery and environmental scatterers, we enforce a minimum effective velocity floor of $v_{\min} = 0.5$ m/s in the Doppler computation. This boundary condition is our approximation and is not part of TR 38.901, which represents scatterer mobility through per-path Doppler components (Clause~7.6.10) \cite{3gpp38901}; it prevents the channel model from degenerating into quasi-static fading when a UE is at rest.

\subsubsection{Stochastic Transient Blockage (CTMC)} \label{sec:blockage} Industrial environments suffer from sudden, severe signal obstructions (e.g., crossing forklifts or moving human workers). We model these transient blockages as a two-state Continuous-Time Markov Chain (CTMC) with state space $\mathcal{S} = \{U, B\}$ (Unblocked, Blocked). The per-UE state $s_u(t) \in \mathcal{S}$ evolves continuously with transition rates $\lambda_{U \to B}$ (blockage onset) and $\lambda_{B \to U}$ (blockage clearance), yielding a steady-state blocking probability of $\pi_B = \lambda_{U \to B} / (\lambda_{U \to B} + \lambda_{B \to U})$ \cite{8643739}. The CTMC represents blockage as an on-off process with exponentially distributed blocked and unblocked intervals, an abstraction used to analyze the impact of mobile blockers in millimeter-wave cellular systems \cite{8643739, 9726709}. Like Model~A of TR 38.901 (Clause~7.6.4.1), it describes blockage stochastically rather than through explicit blocker geometry as in Model~B, but it differs from Model~A, which represents blockers by angular blocking regions around the UE whose positions are spatially and temporally correlated \cite{3gpp38901}.

Consequently, the instantaneous achievable rate $R_u(t)$ for UE $u$ at TTI $t$ is determined by the Adaptive Modulation and Coding (AMC) module from the per-resource-block SINR:

\begin{equation}
    R_u(t) = \mathcal{F}_{AMC} \left( \text{SINR}_u(t) - \zeta_u(t) \right)
    \label{eq:Shannon-capacity-bounded-AMC}
\end{equation}

where $\text{SINR}_u(t)$ is the SINR in decibels [dB] including path loss, shadow fading, and the fading of (\ref{eq:Thechannelcoefficient}), and $\mathcal{F}_{AMC}$ maps it, through the channel quality indicator (CQI) and the block-error-rate curves of Simu5G \cite{9211504}, to the number of bytes that one resource block carries at the selected modulation and coding scheme. $\zeta_u(t) \in \{0, \zeta_B\}$ represents the instantaneous blockage attenuation in decibels [dB]. Specifically, a strictly positive penalty $\zeta_u(t) = \zeta_B > 0$ is applied when the physical channel is obstructed ($s_u(t) = B$), and $\zeta_u(t) = 0$ when the channel is clear ($s_u(t) = U$). Each connection is served at the rate of its UE, so that $R_c(t) = R_{u(c)}(t)$ in the scheduling metrics of Section~\ref{sec:hol-enhanced-lyapunov}..

\subsection{Traffic Model} \label{sec:traffic_model}
The traffic arrivals for each connection $c \in \mathcal{C}$ at TTI $t$ are denoted by $A_c(t)$. This generalized arrival process is strictly demarcated into three distinct operational profiles:

For URLLC connections ($c \in \mathcal{C}_{URLLC}$), $A_c(t)$ represents small, time-critical control and safety messages of the AGV control loop, such as the navigation and safety information that each AGV sends to the industrial controller, characterized by strict periodicity (e.g., 1 ms intervals) and sparse, deterministic payloads (e.g., 100 bytes). In its periodicity, frame size, and deadline requirement, this traffic corresponds to the isochronous traffic type of the TSN profile for industrial automation, which is normally used in control loop tasks and is distinct from the network control traffic, such as time synchronization, that maintains the operation of the network itself \cite{IEC60802}. It likewise corresponds to the isochronous traffic class characterized for TSN over 5G in \cite{9065178}, which is periodic with cycle times below 2 ms and fixed payloads of 30 to 100 bytes. Crucially, every URLLC flow is tightly coupled with a rigid QoS delay budget, $D_{budget, c} \le 10 \text{ ms}$, representing the maximum latency that the control loop tolerates and corresponding to the end-to-end latency bound that TS~22.104 specifies for closed-loop control in process automation \cite{10584103}.

Conversely, for eMBB connections ($c \in \mathcal{C}_{eMBB}$), $A_c(t)$ emulates high-definition industrial video streaming (e.g., automated optical inspection). The packet inter-arrival times are strictly periodic (e.g., 30 fps), but the payload volumes are highly volatile, modeled via a Bernoulli-multiplexed uniform distribution to represent the byte-size difference between video Key-frames (I-frames) and Predictive-frames (P-frames). As in the I/P-frame video traffic model of 3GPP TR~38.838, which is also specified for uplink video \cite{3gpp38838}, each frame is either an I-frame or a P-frame, with I-frames larger than P-frames. The frame type is selected at random, as in the frame-size model of \cite{10.1145/3460797.3460807}, rather than through a fixed group-of-pictures structure, while the uniform size distributions and their parameters, given in Section~\ref{sec:Simulation&Methodology}, are our modelling choice. This generates severe, sudden volume-dominant traffic flows.

Finally, for mMTC connections ($c \in \mathcal{C}_{mMTC}$), $A_c(t)$ represents Industrial Internet of Things (IIoT) sensor telemetry, corresponding to the sporadic best-effort class of \cite{9065178}, which is aperiodic, of low criticality, and of variable payload between 30 and 1500 bytes. Here, massive refers to the large number of connected devices, each transmitting a low volume of delay-tolerant data \cite{10620389}, rather than to the data volume of an individual flow. To capture sporadic, bursty arrivals within that class, these inter-arrival times follow a heavy-tailed Shifted Pareto distribution, a choice consistent with the finding that a generalized Pareto distribution gives the best statistical fit among the tested distributions to measured machine-type traffic \cite{RUIZGUIROLA2025126726}, and the payloads are bimodal. Unlike URLLC, mMTC traffic is fundamentally latency-tolerant, operating under best-effort delivery constraints.

\subsection{Queueing Dynamics} The MAC layer maintains an independent buffer for each logical connection. Let $Q_c(t)$ denote the instantaneous queue backlog (in bits) for connection $c \in \mathcal{C}$, that is, the DRB carrying one traffic class of UE $u(c)$, at the beginning of TTI $t$. Assuming $b_c(t)$ represents the amount of data transmitted during TTI $t$ based on the allocated RBs and $R_c(t)$, the discrete-time queue evolution is given by:

\begin{equation}
Q_c(t+1) = \max \left[ 0, Q_c(t) - b_c(t) \right] + A_c(t)
    \label{eq:discrete-time queue evolution}
\end{equation}

A flow is considered strictly stable if the time-averaged backlog satisfies $\limsup_{T \to \infty} \frac{1}{T} \sum_{t=0}^{T-1} \mathbb{E}[Q_c(t)] < \infty$. However, standard queue stability is insufficient for TSN; the network must also enforce strict temporal bounding.

\section{Proposed HoL-Enhanced Lyapunov Scheduler} \label{sec:hol-enhanced-lyapunov}

Having established the rigorous physical and traffic constraints of the converged 5G-TSN industrial environment, this section details the core algorithmic contribution of our work. We first formalize the deterministic scheduling objective and expose the critical mathematical vulnerabilities of the standard Lyapunov drift-plus-penalty baseline. Subsequently, we introduce the Head-of-Line (HoL) Enhanced framework, detailing the integration of a novel time-aware exponential multiplier designed to mathematically overpower massive eMBB queue volumes so that sparse, deadline-critical flows are not starved by volume dominance.

\subsection{Problem Formulation} In traditional scheduling, a typical objective might be to maximize proportional fairness or aggregate throughput. In converged 5G-TSN networks, the objective paradigm shifts to bounding the worst-case tail latency to guarantee time criticality. We formulate the scheduling problem to minimize the maximum delay experienced by any critical packet, formulated as a min-max optimization:

\begin{equation}
\label{eq:scheduling-min-max}
\begin{split}
    \min \left( \max_{c \in \mathcal{C}_{URLLC}} \omega_c(t) \right) \\
    \text{s.t.} \quad & \omega_c(t) \le D_{budget,c}, \quad \forall c \in \mathcal{C}_{URLLC}
\end{split}
\end{equation}
where $\omega_c(t)$ is the actual waiting time of the packet.

\subsection{The Pure Lyapunov Baseline and its Flaw} Let $\mathbf{Q}(t) = (Q_1(t), Q_2(t), \dots, Q_{|\mathcal{C}|}(t))$ represent the network state vector encompassing all queue backlogs at TTI $t$. The standard Lyapunov framework ensures network-wide stability by defining a scalar quadratic Lyapunov measure of total congestion:
\begin{equation}
L(\mathbf{Q}(t)) = \frac{1}{2} \sum_{c \in \mathcal{C}} Q_c(t)^2
    \label{eq:totalCongestion-scalar quadratic Lyapunov}
\end{equation}

To maintain strict stability, the scheduler must minimize the conditional Lyapunov drift, $\Delta(\mathbf{Q}(t))$, which models the expected change in aggregate congestion over a single TTI transition:

\begin{equation}
\Delta(\mathbf{Q}(t)) = \mathbb{E} \left[ L(\mathbf{Q}(t+1)) - L(\mathbf{Q}(t)) \mid \mathbf{Q}(t) \right]
    \label{eq:conditional-Lyapunov-drift}
\end{equation}

By squaring the queue evolution constraint $Q_c(t+1) \le \max[0, Q_c(t) - b_c(t)] + A_c(t)$, the drift can be upper-bounded. To balance queue stability with network utility (e.g., maximizing throughput or satisfying QoS), the drift-plus-penalty theorem minimizes the bound on $\Delta(\mathbf{Q}(t)) - V \mathbb{E}[U(t) | \mathbf{Q}(t)]$, where $V \ge 0$ is a control parameter and $U(t)$ is the system utility.

Opportunistically minimizing this theoretical upper bound algebraically reduces the MAC-layer scheduling problem to a localized max-weight matching policy at each TTI. Specifically, the base station must allocate RBs to maximize the inner product of the queue vector and the transmission rate vector: $\sum_{c \in \mathcal{C}} Q_c(t) R_c(t)$. To support varying traffic classes, modern QoS-aware Lyapunov variants introduce an exponential tuning parameter \(\alpha > 0\) to shape the queue backlog's influence on the score, alongside a static QoS priority weight \(W_c^\beta\). Values \(\alpha < 1 \)compress the dynamic range of the backlog term, limiting volume dominance, whereas \(\alpha > 1\) amplifies it; \(\alpha = 1\) recovers the classical max-weight policy. This yields the classic scheduling policy that maximizes the following priority score:

\begin{equation}
S_{pure, c}(t) = Q_c(t)^\alpha \cdot R_c(t) \cdot W_c^\beta
    \label{eq:classic-scheduling-policy}
\end{equation}

While the theoretical foundation of this metric robustly stabilizes long-term backlogs across the network, it exhibits a critical vulnerability to heterogeneous traffic in transient environments. Because the metric is fundamentally anchored to $Q_c(t)$, it is volume-dominant. If all connections are ranked by this single score, the continuous, massive volume of $Q_{eMBB}(t)$ mathematically overpowers the scoring equation. The scheduler experiences "threshold blindness," abandoning the URLLC packet in the buffer because its small TSN volume cannot mathematically compete with massive eMBB backlogs. Per-class DRBs with strict URLLC precedence, as in the class-isolated hybrid scheduler of~\cite{11555379}, remove this inter-class failure but not the volume dominance among the URLLC connections of different AGVs: a backlog-driven score still serves the AGV with the larger URLLC backlog before the one whose oldest packet is closest to its deadline. This forces the critical control packet to violate its $D_{budget,c}$, causing system failure.

\subsection{Head-of-Line (HoL) Delay Estimator} 
To cure this temporal blindness, we introduce a Head-of-Line (HoL) age estimator directly into the MAC-layer control loop. For any active queue $Q_c(t) > 0$, the scheduler records the exact arrival timestamp $t_{arr, c}^*$ of the oldest packet currently awaiting transmission. The instantaneous HoL delay is computed as:

\begin{equation}
D_{HoL, c}(t) = t - t_{arr, c}^*
    \label{eq:instantaneous-HoL-delay}
\end{equation}

This transforms the scheduler from being strictly queue-aware (volume-based) to explicitly deadline-aware (time-based).

\subsection{The Exponential Age Penalty and Isolation Multiplier} 
To mathematically overpower the dominance of eMBB volume, we propose a bounded exponential age-penalty multiplier. To prevent numerical instability from unbounded head-of-line growth while maintaining sufficient sensitivity to approaching deadlines, we first apply a clipping parameter \(\eta\) to establish a bounded effective age:

\begin{equation}
D_{bounded, c}(t) = \min \left( D_{HoL, c}(t), \eta \cdot D_{budget, c} \right)
    \label{eq:clipping-parameter}
\end{equation}

where \(\eta = 5.0\) is the largest integer for which the maximum multiplier $\rho$ (defined below) stays within the URLLC isolation bound $\Gamma_c = 10^{12}$; in our simulations, the 99th-percentile URLLC delay stays below $5 D_{budget,c}$ up to $N = 10$ for all schedulers except BestFit.

We then construct the HoL multiplier \(M_{HoL, c}(t)\) using an exponential growth function scaled by a tuning parameter \(\gamma\):

\begin{equation}
M_{HoL, c}(t) = \exp \left( \gamma \cdot \frac{D_{bounded, c}(t)}{D_{budget, c}} \right)
    \label{eq:HoL multiplier}
\end{equation}

where \(\gamma\) sets the growth rate of the multiplier and \(\rho = \exp(\gamma \cdot \eta)\) is the resulting maximum multiplier effect, attained when the head-of-line delay reaches the clipping bound, i.e. \(M_{HoL, c}(\eta \cdot D_{budget, c}) = \rho\). We use \(\gamma = 5.0\), the smallest integer for which a flow whose head-of-line packet has reached its budget (\(e^{\gamma} \approx 148\)) outranks a fresh flow with the largest weight and channel advantage among the non-critical classes (\(W_{eMBB}/W_{mMTC} \cdot R_{\max}/R_{\min} \approx 4 \times 24\)), which with \(\eta = 5.0\) yields \(\rho \approx 7.2 \times 10^{10}\). The multiplier is therefore bounded, but the bound is enforced by the clipping parameter \(\eta\) rather than by a small value of \(\rho\): the design intent is a steep, strictly monotonic urgency gradient that separates packets by deadline proximity across the full range of delays observed under congestion, with clipping guaranteeing numerical stability during prolonged blockages.

Finally, to resolve in-class competition among multiple TSN endpoints and ensure absolute preemption over eMBB payloads, we apply a strict URLLC priority isolation booster, \((\Gamma_c)\), where \((\Gamma_c = 10^{12})\) if \((D_{budget, c} \le 10\text{ ms})\), and \((\Gamma_c = 1)\) otherwise. The value is determined structurally rather than empirically: it must exceed the maximum possible ratio of a non-critical score to a critical score, \(((\rho \cdot R_{\max} \cdot W_{eMBB}^\beta) / (R_{\min} \cdot W_{URLLC}^\beta))\), ensuring that any URLLC flow always receives higher priority than non-critical flows regardless of their instantaneous scores. We select \(10^{12}\) as a conservative bound satisfying this condition within double-precision arithmetic, ensuring that a single URLLC packet nearing its deadline preempts any concurrent eMBB or mMTC demand regardless of channel quality or class weight. This is about an order of magnitude above the required ratio (\(\approx 8.5 \times 10^{10}\)); because \(\Gamma_c\) scales every URLLC score equally, any larger value yields the same scheduling decisions, whereas an infinite value would make all URLLC scores tie and remove their head-of-line ordering.

The proposed HoL-Enhanced Lyapunov scoring function becomes:

\begin{equation}
S_{HoL, c}(t) = M_{HoL, c}(t) \cdot R_c(t) \cdot W_c^\beta \cdot \Gamma_c
    \label{eq:HoL-Enhanced-Lyapunov-scoring}
\end{equation}

The exponential form is chosen because it provides a smooth, monotonically increasing function that remains bounded (due to clipping) while providing sufficient incentive to prioritize URLLC packets as they approach their deadline. Unlike polynomial functions, the exponential ensures that the priority increase accelerates as the deadline approaches, providing stronger urgency signaling near the deadline threshold. This characteristic better matches the increasing criticality of URLLC packets as their deadline nears.

\subsection{Algorithmic Complexity and Feasibility} 
The computational viability of the proposed framework is a critical necessity given the rigid 0.5 ms TTI duration of industrial 5G profiles. Because $D_{HoL, c}(t)$ requires only a simple local timestamp lookup for each active queue, the state extraction phase requires $O(|\mathcal{C}|)$ operations, where $|\mathcal{C}|$ is the number of active logical connections with $Q_c(t) > 0$. The evaluation of the priority score $S_{HoL, c}(t)$ and the subsequent sorting of the candidate list scales with $O(|\mathcal{C}| \log |\mathcal{C}|)$. Since $|\mathcal{C}|$ is dimensionally small within a single localized cell, the computational overhead is highly deterministic and negligible, allowing the HoL-Enhanced algorithm to execute seamlessly within the baseband unit's processing window without necessitating predictive machine learning overhead. 

Furthermore, to ensure numerical stability when transitioning from high-precision software simulation to hardware-based fixed-point implementations common in 5G baseband processors, we recommend scaling all scoring functions by a common normalization factor prior to the scheduling comparison. Specifically, the maximum theoretical value of the proposed scoring function is strictly bounded by $S_{\max} = \rho \cdot R_{\max} \cdot W_{\max}^\beta \cdot \Gamma_{\max}$, where $R_{\max}$ and $W_{\max}$ represent the system-defined maximum limits for achievable spectral efficiency and QoS priority weight, $\rho$ bounds the age-penalty multiplier, and $\Gamma_{\max} = 10^{12}$ is the isolation booster applied to the critical class. By scaling the instantaneous score by this deterministic upper bound ($S_{HoL,c}(t) / S_{\max}$), the scheduling metric is cleanly normalized to a $[0, 1]$ range. This arithmetic scaling ensures robust numerical precision and prevents register overflow within standard fixed-point Arithmetic Logic Units (ALUs), all without altering the strict ordinal ranking of the scheduled connections.

Fig. \ref{fig:tsn-scheduler} illustrates the structural flow and interaction of the proposed scheduling algorithm. At the input stage, the framework continuously monitors the independent Head-of-Line waiting times (e.g., $D_{HoL, 1}(t)$ and $D_{HoL, n}(t)$) for the URLLC connections of multiple competing UEs, alongside instantaneous channel conditions ($R_c(t)$) and massive eMBB queue backlogs. The processing engine evaluates every connection with the single scoring function of (\ref{eq:HoL-Enhanced-Lyapunov-scoring}), so that the exponential age-penalty multiplier is applied to URLLC, eMBB, and mMTC alike, each normalized by its own delay budget. The classes are separated within that one expression by the isolation booster $\Gamma_c$ alone, which takes the value $10^{12}$ for the connections whose budget marks them as critical and unity for all others. Because $\Gamma_c$ is chosen to exceed the largest attainable ratio between a non-critical and a critical score, this single metric induces a strict ordering between the classes without any branch in the scoring logic, and the tiering visible in the output allocation grid follows from that expression rather than from a separate classification stage. It allows the framework to fully preempt non-critical payloads while dynamically resolving in-class URLLC resource collisions based on real-time packet urgency.

Algorithm \ref{alg:hol_lyapunov} summarizes the operational logic of the proposed HoL-Enhanced Lyapunov scheduler. Several practical advantages emerge from this delay-aware MAC-layer design. First, the computational complexity is strictly minimized, scaling linearly with the number of active user connections ($O(|\mathcal{C}|)$) rather than requiring complex combinatorial permutations of Resource Block (RB) allocations. Because the number of active queues evaluated during any single decision phase is bounded, the HoL tracking and scoring logic introduces negligible processing overhead, guaranteeing execution strictly within the 0.5 ms TTI window. Furthermore, because the control law is entirely deterministic and driven by instantaneous feedback, such as real-time queue age and current spectral efficiency, the system remains inherently robust to fast fading, shadowing, and transient blockages without necessitating computationally expensive channel prediction mechanisms or training-heavy machine learning models. Since the channel enters the score only through the reported spectral efficiency, no assumption on the fading distribution is made, and the Rician fading of Section~\ref{sec:system_model} is the case evaluated here.

As detailed in Algorithm \ref{alg:hol_lyapunov}, the priority scores for all active queues are evaluated independently prior to the physical resource allocation phase. While a strictly coupled formulation might attempt to iteratively distribute fractional RBs among competing flows, the proposed approach simplifies this by requesting an unbounded allocation, allowing the AMC module to natively clip the grant to the exact queue depth. Additionally, because the scheduler replaces the backlog term with the head-of-line urgency multiplier, so massive eMBB queue volumes cannot dominate the score, and isolates URLLC through the $\Gamma_c = 10^{12}$ multiplier, it avoids the need for heavy cross-flow orchestration. Cell-wide congestion and physical layer blockages are inherently reflected in the exponential growth of the HoL timer ($D_{HoL}$) and the instantaneous achievable rate ($R_c$). Therefore, executing a localized scoring equation preserves the algorithm's computational simplicity without sacrificing the optimal deterministic latency boundaries required for TSN.

\begin{algorithm}
\caption{Proposed HoL-Enhanced Lyapunov Scheduling Framework}
\label{alg:hol_lyapunov}
\begin{algorithmic}[1]
\Require Tuning parameters $\alpha, \beta, \gamma$
\Require Set of active user connections $\mathcal{C}$
\Require System-defined maximum score bound $S_{\max}$

\Loop { for each TTI $t$}

    \State Initialize priority candidate list $\mathcal{L} \gets \emptyset$
    
    \State \textbf{// Phase 1: Delay-Aware Scoring Formulation}
    \ForAll {connection $c \in \mathcal{C}$ with backlog $Q_c(t) > 0$}
        
        \State \textbf{// Retrieve instantaneous parameters}
        \State $D_{budget} \gets \text{Target QoS Delay Bound for } c$
        \State $D_{HoL} \gets \text{Current waiting time of oldest}$
\Statex \hspace*{1.5cm} $\text{packet in } Q_c(t)$

        \State $R_c(t) \gets \text{Achievable spectral efficiency}$
\Statex \hspace*{1.5cm} $\text{from AMC module}$
        
        \State $W_c \gets \text{Static QoS weight for } c$
        
        \State \textbf{// Prevent Threshold Blindness via upper bounding}
        \State $D_{bounded} \gets \min(D_{HoL}, D_{budget} \times 5.0)$ 
        
        \State \textbf{// Compute HoL-Enhanced Drift-plus-Penalty}
        \State $M_{HoL} \gets \exp\left(\gamma \cdot \frac{D_{bounded}}{D_{budget}}\right)$
        \State $S_c \gets M_{HoL} \cdot R_c(t) \cdot W_c^\beta$
        
        \State \textbf{// Apply Strict Priority Isolation for TSN Traffic}
        \If {$D_{budget} \le 10 \text{ ms}$}
            \State $S_c \gets S_c \times 10^{12}$
        \EndIf
        
        \State \textbf{// Normalize to prevent fixed-point ALU overflow}
        \State $S_c \gets S_c / S_{\max}$
        
        \State Append $\{c, S_c\}$ to $\mathcal{L}$
    \EndFor

    \State \textbf{// Phase 2: Physical Resource Allocation}
    \State Sort $\mathcal{L}$ in descending order of priority score $S_c$
    \State $RBs_{available} \gets \text{Total capacity for TTI } t$
    
    \ForAll {candidate $c^* \in \mathcal{L}$}
        \If {$RBs_{available} == 0$}
            \State \textbf{break}
        \EndIf
        \State $RBs_{required} \gets \text{Calculate RBs for } Q_{c^*}(t) \text{ based on } R_{c^*}(t)$
        \State $RBs_{granted} \gets \min(RBs_{required}, RBs_{available})$
        \State $\text{Allocate } RBs_{granted} \text{ to } c^*$
        \State $RBs_{available} \gets RBs_{available} - RBs_{granted}$
    \EndFor

\EndLoop
\end{algorithmic}
\end{algorithm}

\section{Simulation Environment and Methodology}
\label{sec:Simulation&Methodology}
To evaluate the proposed HoL-Enhanced Hybrid Lyapunov scheduler under realistic industrial conditions, a comprehensive discrete-event simulation campaign was conducted. The methodology follows 3GPP channel model guidelines for Industrial Factory (InF) environments and incorporates the custom physical-layer channel physics and QoS-flow architecture detailed in the preceding sections. All results reported in Section~\ref{sec:Results} are averaged over 10 independent simulation runs with 95\% confidence intervals.

\subsection{Simulation Framework: OMNeT++ and Simu5G}
The evaluation framework is built upon the OMNeT++~6.4 discrete-event simulation engine with the INET~4.6 protocol suite, utilizing the Simu5G framework 1.4.1 SDAP version \cite{9211504, 11229920} to emulate the 5G New Radio (NR) protocol stack. Simu5G provides a granular, standard-compliant implementation of the 5G user plane, including the MAC and Radio Link Control (RLC) layers. The Simu5G version used in this work includes an SDAP layer with QoS-Flow-to-DRB mapping, allowing each traffic class to be carried over its own Data Radio Bearer (DRB) with an individual 5QI-style QoS profile, including delay budget, packet error rate target, GBR configuration, and priority. For this study, the native Simu5G source code was substantially extended in two directions. First, the proposed Hybrid Lyapunov and HoL-Enhanced Hybrid Lyapunov algorithms were implemented as pluggable scheduling policies within the gNB MAC scheduler framework, selectable per run alongside the unmodified stock policies used as baselines. Second, the physical-layer channel modules were extended to inject the true complex I/Q Rician fading model, the AR-1 spatial-consistency filter, and the continuous Doppler phase-rotation logic. This architectural approach ensures that high-level scheduling decisions are evaluated against sub-millisecond, physically accurate radio-frequency propagation mechanics rather than abstracted statistical success rates.

\subsection{Scenario Configuration: Industrial Factory (InF)} The simulation geometry emulates a dense-clutter Industrial Factory (InF-DL) environment as per 3GPP TR~38.901 \cite{3gpp38901}. The factory floor spans a 150\,m~$\times$~100\,m area, served by a single gNB mounted centrally at a height of 8\,m. The network operates in the C-Band at a carrier frequency of $f_c = 3.8$\,GHz with numerology~$\mu = 1$ (30\,kHz subcarrier spacing), which maps to a 0.5\,ms TTI. The cell is provisioned with a fixed budget of 51 physical resource blocks (20\,MHz), representative of a dedicated private-5G industrial allocation; UE and gNB transmit powers are 23 and 24\,dBm respectively, and link adaptation targets a block-error rate of $10^{-2}$. On top of the InF-DL path-loss model, the channel applies the spatially correlated Rician fading process of Section~\ref{sec:system_model} and stochastic clutter blockage with a 20\,dB attenuation penalty. 

Mobile robotic nodes, representing AGVs and mobile sensor platforms with a 0.5\,m antenna height, navigate the factory floor using a random-waypoint mobility model. To exercise the Doppler-induced fading effects described in Section~\ref{sec:system_model}, AGV velocities are uniformly distributed $v \sim \mathrm{Unif}[0.5, 3.0]$\,m/s. The strict minimum velocity floor of $v_{\min} = 0.5$\,m/s is maintained to emulate the ambient Doppler spread of moving environmental scatterers, preventing the spatially correlated fading process from decaying into an unrealistic quasi-static state during temporary AGV halts. Traffic flows uplink from the AGVs through the gNB and across a 1\,Gb/s TSN bridge to an industrial controller end-station, where end-to-end latency is measured at the application sink.

\subsection{Traffic Profiling and QoS Parameters} Each AGV concurrently generates three flows, mapped by the SDAP layer onto three dedicated DRBs. To prevent artificial synchronization artifacts, all application start phases are uniformly staggered with production offsets $t_{\mathrm{offset}} \sim \mathrm{Unif}[0, 2]$\,s. 

\begin{itemize} 
    \item \textbf{URLLC / TSN control flow} (DRB~0; GBR, priority~1, $D_{\mathrm{budget}} = 10$\,ms, PER target $10^{-5}$): representing the periodic navigation and safety messages that each AGV sends to the industrial controller in its control loop (Section~\ref{sec:traffic_model}), this traffic is strictly deterministic: 100\,B payloads injected with a rigid 1\,ms periodicity (0.8\,Mb/s per AGV). 
    
    \item \textbf{eMBB video flow} (DRB~1; GBR, priority~2, $D_{\mathrm{budget}} = 100$\,ms, PER target $10^{-3}$): representing continuous industrial video surveillance (30\,fps automated optical inspection), generating $\approx 3.9$\,Mb/s per node. Rather than generic heavy-tailed bursts, payload generation follows the I/P-frame structure of Section~\ref{sec:traffic_model} via a Bernoulli-multiplexed uniform distribution: every 33.333\,ms the application emits either a Key-frame (I-frame) drawn from $\mathrm{Unif}[50, 60]$\,KB with probability 0.05, or a Predictive-frame (P-frame) drawn from $\mathrm{Unif}[12.4, 16.4]$\,KB with probability 0.95. These values are our choice; they give an I/P size ratio of about 3.8, above the ratios of 1.5 to 3 used in TR~38.838~\cite{3gpp38838}, so the stream is burstier than the 3GPP reference. This periodic but highly volatile payload spiking is explicitly designed to flood the UE uplink buffers and provoke cross-class starvation.
    
    \item \textbf{mMTC telemetry flow} (DRB~2; non-GBR, priority~3, $D_{\mathrm{budget}} = 300$\,ms, PER target $10^{-2}$): representing massive background IIoT chatter with inter-arrival times following a shifted Pareto distribution (shape $\kappa = 1.2$, scale $x_m = 50$\,ms, shift 20\,ms) and a bimodal payload: a 64\,B telemetry heartbeat with a 0.3 probability of appending an extended 1386\,B diagnostic block ($\approx 17$\,kb/s per node). \end{itemize} 
    
This asymmetric traffic mix, the eMBB flow carries roughly $80\%$ of the offered bytes while the URLLC flow carries the deadline-critical control loop ensures that the proposed scheduler's strict isolation booster ($\Gamma_c = 10^{12}$) and fixed-point normalization logic are aggressively stress-tested against the volatile I-frame payload spikes of the concurrent eMBB streams.

\subsection{Experimental Design and Compared Schedulers}
The independent variable of the study is the fleet size $N$, swept over $N \in \{5, 10, 11, 12, 13, 14, 15, 20, 25, 30\}$ on the fixed 51 physical RB budget; the fine 1-AGV resolution between $N = 10$ and $N = 15$ brackets the cell's capacity boundary, while $N = 20$--$30$ probes controlled overload up to $2.5\times$ the supported load. Every (scheduler, $N$) operating point is simulated for 10 independent runs with separate random seeds, of 300\,s each (the initial 20\,s transient period is discarded), for a total of 700 production runs. The two proposed schedulers: the HoL-Enhanced Hybrid Lyapunov scheduler (HoL-Lyapunov) and its Hybrid-Strict Priority based Lyapunov counterpart (Hybrid Lyapunov) \cite{11555379}, retained as an ablation of the head-of-line delay term are benchmarked against five scheduling policies from the stock Simu5G framework and State-of-the-art: maximum channel-quality scheduling (Max~C/I), Proportional Fair, and the best-fit resource allocator (BestFit), all three distributed with Simu5G \cite{9211504}, together with QoS-aware Proportional Fair (QoS-PF) \cite{11309032} and the delay-aware DQoS scheduler \cite{10333533}. The plain Hybrid Lyapunov variant is evaluated with the tuning values of its original formulation \cite{11555379} \(\alpha = 1.2\), \(\beta = 1.0\), held fixed across the entire load range; the proposed HoL-Enhanced variant uses the same \((\beta)\) and \(\gamma = 5.0\), \(\eta = 5.0\). Because the head-of-line multiplier replaces the backlog term outright, \((\alpha)\) does not enter the proposed scheduler's score, so the ablation isolates the substitution of the exponential urgency term for the linear backlog term, with all remaining components like achievable rate, QoS weight, and the isolation booster \((\Gamma_c)\) identical in both variants.

Across all runs, the schedulers are compared on the following metrics. Latency is measured end-to-end (application source in the AGV to application sink behind the TSN bridge) and recorded per class as a fine-grained histogram, from which all percentile and deadline statistics are derived. The analysis focuses on the following metrics: \begin{enumerate} 
    \item \textbf{Packet Delivery Ratio (PDR):} delivered packets over generated packets, per class. Under saturation the radio bearers shed load, making PDR the primary indicator for which classes a scheduler protects. 
    
    \item \textbf{Deadline-miss ratio:} the fraction of \emph{delivered} packets exceeding the class delay budget (10 / 100 / 300\,ms), evaluated from the latency histogram mass above the budget. For URLLC, we also perform an additional analysis after the simulations using relaxed end-to-end latency thresholds of 15, 20, and 30\,ms, because the architecture imposes a minimum latency limit discussed in Section~\ref{sec:Results}. 
    
    \item \textbf{QoS satisfaction (headline metric):} the percentage of all generated packets that are successfully delivered before their deadline, calculated as $\mathrm{QoS} = \mathrm{PDR} \times (1 - \mathrm{miss})$. Unlike average latency measured only on delivered packets, this metric avoids bias toward packets that were not dropped. A scheduler cannot improve its QoS score by discarding late packets; it must deliver more packets within the required deadline.
    
    \item \textbf{Tail latency:} the 99th-percentile end-to-end latency, obtained by linear interpolation inside histogram bins. Given the zero-tolerance nature of industrial control loops, tail behavior is weighted more heavily than the mean; mean latency and delay variation (jitter, reported as the standard deviation of the latency distribution) are reported alongside for completeness. 
    
    \item \textbf{Goodput and resource-block utilization:} per-class delivered throughput, and the fraction of the 51-PRB uplink budget actually scheduled. The latter verifies that observed degradation is a genuine capacity limit shared by all schedulers rather than a configuration artifact. \end{enumerate} 
    
Mean latency is interpreted as a performance measure only when $\mathrm{PDR} \approx 100\%$; under saturation it is reported solely to expose the survivorship bias it carries (Section~\ref{sec:overload}).

\section{Performance Evaluation and Results}
\label{sec:Results}
\begin{figure*}[t]
\centering
\includegraphics[width=\textwidth]{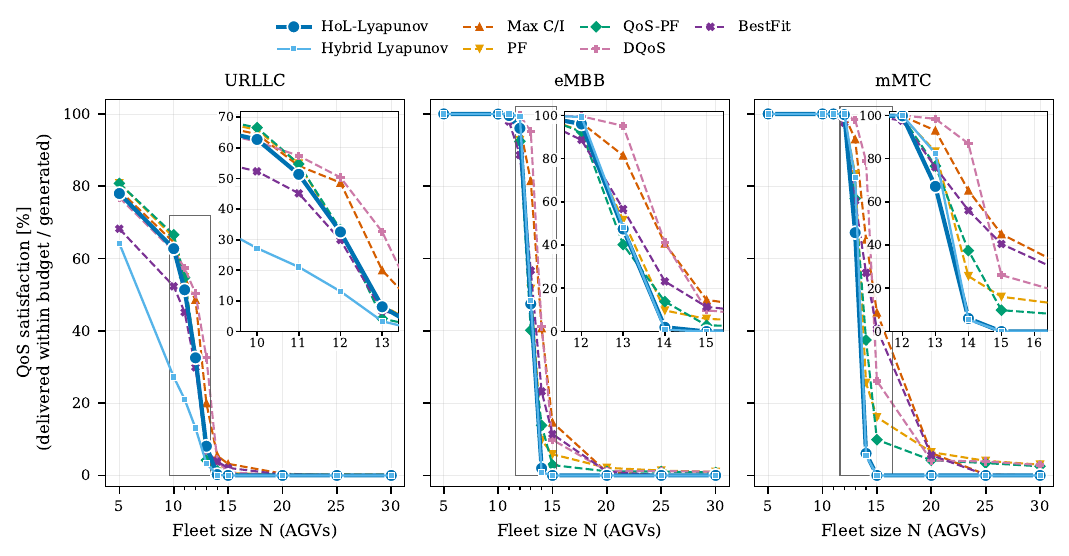}
\caption{QoS satisfaction (packets delivered within the class delay budget, as a fraction of packets generated) vs.\ fleet size $N$, per traffic class. Markers: mean over 10 replications. Solid lines: proposed schedulers; dashed: baselines. The inset in each panel expands the load range marked by the rectangle, over which the schedulers separate.}
\label{fig:qos_sat}
\end{figure*}

\begin{figure*}[t]
\centering
\includegraphics[width=\textwidth]{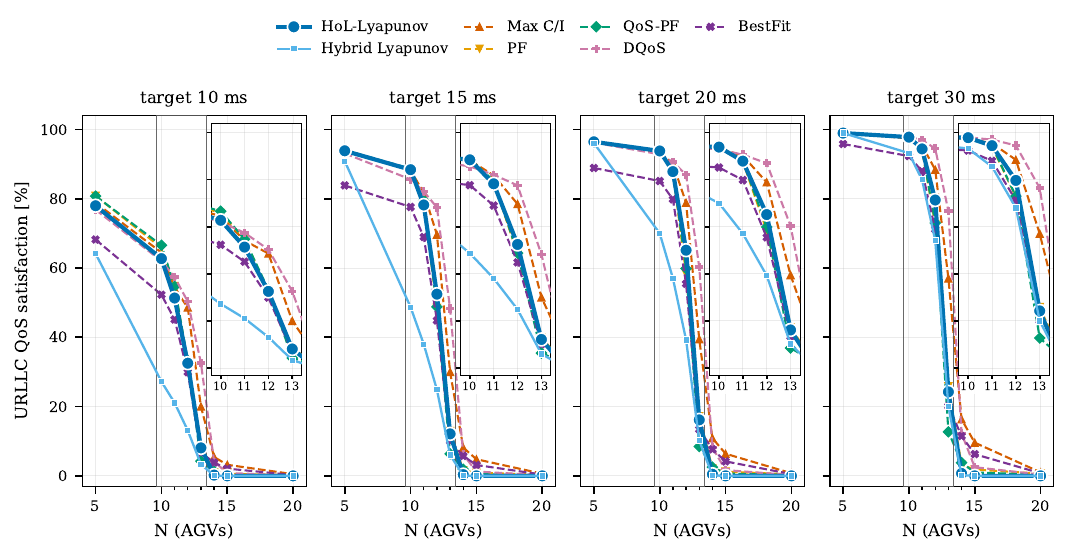}
\caption{URLLC QoS satisfaction vs.\ fleet size $N$ for post-hoc end-to-end latency targets of 10/15/20/30 ms (the deployment's budget and three relaxed targets). Markers: mean over 10 replications. The axis is truncated at $N = 20$; at the two omitted points ($N = 25$ and $N = 30$) no scheduler exceeds 0.6\% at any of the four targets. The inset in each panel expands the $N = 10$ to $13$ band marked on the axes, over which the schedulers separate.}
\label{fig:urllc_targets}
\end{figure*}

The evaluation resolves two operating regimes with qualitatively different scheduler behavior: a feasible-load regime ($N \le 12$), in which every scheduler delivers essentially all traffic and the contest is over deadline compliance, and an overload regime ($N \ge 14$), in which no scheduler can meet URLLC deadlines and the contest is over what is preserved. Figures~\ref{fig:qos_sat}--\ref{fig:urllc_targets} and Tables~\ref{tab:nominal}--\ref{tab:overload} report means over 10 replications.

\begin{table*}[t]
\centering
\caption{Per-class results at the nominal operating point ($N=12$ AGVs); mean $\pm$ 95\% CI over 10 replications. Latency statistics cover delivered packets only.}
\label{tab:nominal}
\begin{tabular}{llrrrrrrr}
\toprule
Scheduler & Class & PDR [\%] & QoS-sat [\%] & Mean lat. [ms] & Jitter [ms] & p99 [ms] & Miss [\%] & Goodput [Mb/s] \\
\midrule
HoL-Lyapunov & URLLC & 99.99 $\pm$ 0.00 & 32.53 $\pm$ 1.81 & 24.0 $\pm$ 2.8 & 34.1 $\pm$ 6.6 & 167 $\pm$ 28 & 67.5 $\pm$ 1.8 & 9.60 $\pm$ 0.00 \\
 & eMBB & 99.99 $\pm$ 0.01 & 96.03 $\pm$ 1.19 & 35.2 $\pm$ 3.0 & 34.6 $\pm$ 6.6 & 180 $\pm$ 28 & 4.0 $\pm$ 1.2 & 47.27 $\pm$ 0.01 \\
 & mMTC & 100.00 $\pm$ 0.01 & 99.80 $\pm$ 0.12 & 24.3 $\pm$ 2.6 & 32.7 $\pm$ 5.1 & 168 $\pm$ 26 & 0.2 $\pm$ 0.1 & 0.20 $\pm$ 0.01 \\
Hybrid Lyapunov & URLLC & 100.00 $\pm$ 0.00 & 13.11 $\pm$ 0.85 & 26.0 $\pm$ 1.2 & 16.1 $\pm$ 1.4 & 82 $\pm$ 9 & 86.9 $\pm$ 0.9 & 9.60 $\pm$ 0.00 \\
 & eMBB & 100.00 $\pm$ 0.00 & 99.29 $\pm$ 0.32 & 35.7 $\pm$ 1.3 & 17.1 $\pm$ 1.4 & 92 $\pm$ 10 & 0.7 $\pm$ 0.3 & 47.29 $\pm$ 0.03 \\
 & mMTC & 100.00 $\pm$ 0.01 & 100.00 $\pm$ 0.01 & 26.1 $\pm$ 1.3 & 16.0 $\pm$ 1.3 & 81 $\pm$ 9 & 0.0 $\pm$ 0.0 & 0.20 $\pm$ 0.01 \\
Max C/I & URLLC & 100.00 $\pm$ 0.00 & 48.61 $\pm$ 2.77 & 19.1 $\pm$ 0.9 & 35.2 $\pm$ 2.7 & 190 $\pm$ 14 & 51.4 $\pm$ 2.8 & 9.60 $\pm$ 0.00 \\
 & eMBB & 100.00 $\pm$ 0.00 & 96.57 $\pm$ 0.39 & 26.3 $\pm$ 0.9 & 36.6 $\pm$ 2.7 & 203 $\pm$ 14 & 3.4 $\pm$ 0.4 & 47.33 $\pm$ 0.03 \\
 & mMTC & 100.00 $\pm$ 0.01 & 99.68 $\pm$ 0.08 & 19.6 $\pm$ 1.0 & 35.9 $\pm$ 3.3 & 192 $\pm$ 17 & 0.3 $\pm$ 0.1 & 0.20 $\pm$ 0.01 \\
PF & URLLC & 99.99 $\pm$ 0.01 & 32.26 $\pm$ 1.66 & 30.9 $\pm$ 1.6 & 50.6 $\pm$ 3.8 & 262 $\pm$ 17 & 67.7 $\pm$ 1.7 & 9.60 $\pm$ 0.00 \\
 & eMBB & 99.99 $\pm$ 0.01 & 92.76 $\pm$ 0.66 & 41.8 $\pm$ 1.7 & 51.3 $\pm$ 3.8 & 275 $\pm$ 17 & 7.2 $\pm$ 0.7 & 47.31 $\pm$ 0.06 \\
 & mMTC & 99.99 $\pm$ 0.01 & 99.42 $\pm$ 0.15 & 30.6 $\pm$ 1.6 & 48.2 $\pm$ 4.3 & 251 $\pm$ 18 & 0.6 $\pm$ 0.2 & 0.21 $\pm$ 0.01 \\
QoS-PF & URLLC & 99.99 $\pm$ 0.01 & 33.27 $\pm$ 1.76 & 31.8 $\pm$ 3.3 & 55.0 $\pm$ 6.9 & 284 $\pm$ 37 & 66.7 $\pm$ 1.8 & 9.60 $\pm$ 0.00 \\
 & eMBB & 99.98 $\pm$ 0.01 & 92.43 $\pm$ 1.36 & 42.6 $\pm$ 3.3 & 55.7 $\pm$ 6.9 & 297 $\pm$ 37 & 7.5 $\pm$ 1.4 & 47.28 $\pm$ 0.06 \\
 & mMTC & 99.98 $\pm$ 0.01 & 99.14 $\pm$ 0.29 & 31.4 $\pm$ 3.6 & 52.6 $\pm$ 8.0 & 279 $\pm$ 35 & 0.8 $\pm$ 0.3 & 0.20 $\pm$ 0.01 \\
DQoS & URLLC & 100.00 $\pm$ 0.00 & 50.33 $\pm$ 1.40 & 12.9 $\pm$ 0.1 & 11.9 $\pm$ 0.6 & 62 $\pm$ 2 & 49.7 $\pm$ 1.4 & 9.60 $\pm$ 0.00 \\
 & eMBB & 100.00 $\pm$ 0.00 & 99.59 $\pm$ 0.06 & 17.1 $\pm$ 0.2 & 13.8 $\pm$ 0.6 & 74 $\pm$ 2 & 0.4 $\pm$ 0.1 & 47.27 $\pm$ 0.03 \\
 & mMTC & 100.00 $\pm$ 0.01 & 99.99 $\pm$ 0.01 & 13.2 $\pm$ 0.2 & 11.9 $\pm$ 1.2 & 62 $\pm$ 5 & 0.0 $\pm$ 0.0 & 0.20 $\pm$ 0.01 \\
BestFit & URLLC & 100.00 $\pm$ 0.01 & 29.90 $\pm$ 4.11 & 43.1 $\pm$ 2.3 & 87.7 $\pm$ 6.6 & 449 $\pm$ 34 & 70.1 $\pm$ 4.1 & 9.60 $\pm$ 0.00 \\
 & eMBB & 99.99 $\pm$ 0.01 & 88.65 $\pm$ 0.70 & 52.2 $\pm$ 2.4 & 89.1 $\pm$ 6.7 & 461 $\pm$ 34 & 11.3 $\pm$ 0.7 & 47.35 $\pm$ 0.04 \\
 & mMTC & 100.00 $\pm$ 0.01 & 97.56 $\pm$ 0.44 & 42.0 $\pm$ 3.3 & 83.9 $\pm$ 8.9 & 438 $\pm$ 42 & 2.4 $\pm$ 0.4 & 0.20 $\pm$ 0.02 \\
\bottomrule
\end{tabular}
\end{table*}

\begin{table*}[t]
\centering
\caption{Overload behavior at $N=30$ AGVs: per-class degradation mode; mean $\pm$ 95\% CI over 10 replications.}
\label{tab:overload}
\begin{tabular}{llrrr}
\toprule
Scheduler & Class & PDR [\%] & QoS-sat [\%] & Goodput [Mb/s] \\
\midrule
HoL-Lyapunov & URLLC & 82.7 $\pm$ 0.1 & 0.00 $\pm$ 0.00 & 19.85 $\pm$ 0.02 \\
 & eMBB & 0.0 $\pm$ 0.0 & 0.00 $\pm$ 0.00 & 0.00 $\pm$ 0.00 \\
 & mMTC & 82.5 $\pm$ 0.3 & 0.00 $\pm$ 0.00 & 0.37 $\pm$ 0.01 \\
Hybrid Lyapunov & URLLC & 74.1 $\pm$ 0.6 & 0.00 $\pm$ 0.00 & 17.79 $\pm$ 0.13 \\
 & eMBB & 0.4 $\pm$ 0.0 & 0.00 $\pm$ 0.00 & 0.40 $\pm$ 0.05 \\
 & mMTC & 75.0 $\pm$ 0.6 & 0.00 $\pm$ 0.00 & 0.34 $\pm$ 0.01 \\
Max C/I & URLLC & 65.7 $\pm$ 0.1 & 0.02 $\pm$ 0.05 & 15.76 $\pm$ 0.03 \\
 & eMBB & 20.0 $\pm$ 0.2 & 0.02 $\pm$ 0.03 & 20.96 $\pm$ 0.25 \\
 & mMTC & 68.6 $\pm$ 0.7 & 0.01 $\pm$ 0.02 & 0.30 $\pm$ 0.01 \\
PF & URLLC & 46.4 $\pm$ 0.1 & 0.26 $\pm$ 0.35 & 11.13 $\pm$ 0.02 \\
 & eMBB & 44.5 $\pm$ 0.1 & 0.97 $\pm$ 0.03 & 52.39 $\pm$ 0.08 \\
 & mMTC & 49.0 $\pm$ 1.2 & 2.98 $\pm$ 0.17 & 0.23 $\pm$ 0.01 \\
QoS-PF & URLLC & 44.4 $\pm$ 0.1 & 0.08 $\pm$ 0.01 & 10.66 $\pm$ 0.03 \\
 & eMBB & 43.8 $\pm$ 0.1 & 0.81 $\pm$ 0.12 & 51.74 $\pm$ 0.13 \\
 & mMTC & 46.4 $\pm$ 1.0 & 2.53 $\pm$ 0.33 & 0.23 $\pm$ 0.01 \\
DQoS & URLLC & 45.6 $\pm$ 0.2 & 0.10 $\pm$ 0.01 & 10.95 $\pm$ 0.04 \\
 & eMBB & 44.5 $\pm$ 0.1 & 1.02 $\pm$ 0.10 & 52.53 $\pm$ 0.08 \\
 & mMTC & 45.9 $\pm$ 1.3 & 2.87 $\pm$ 0.38 & 0.23 $\pm$ 0.01 \\
BestFit & URLLC & 64.2 $\pm$ 0.2 & 0.00 $\pm$ 0.00 & 15.41 $\pm$ 0.06 \\
 & eMBB & 19.0 $\pm$ 0.3 & 0.00 $\pm$ 0.01 & 19.85 $\pm$ 0.40 \\
 & mMTC & 67.5 $\pm$ 0.5 & 0.00 $\pm$ 0.00 & 0.29 $\pm$ 0.01 \\
\bottomrule
\end{tabular}
\end{table*}

\subsection{Capacity threshold}
\label{sec:capacity_threshold}
\begin{figure*}[t]
\centering
\includegraphics[width=\textwidth]{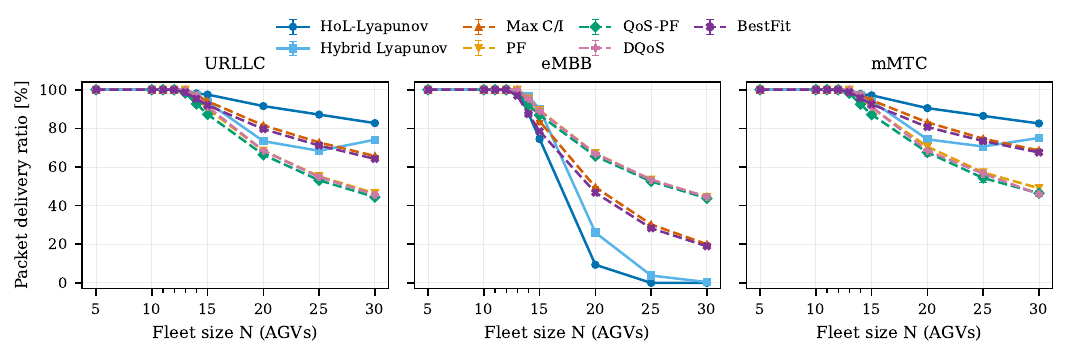}
\caption{Packet delivery ratio vs.\ fleet size $N$, per traffic class (mean $\pm$ 95\% CI over 10 replications).}
\label{fig:pdr}
\end{figure*}

\begin{figure}[t]
\centering
\includegraphics[width=\columnwidth]{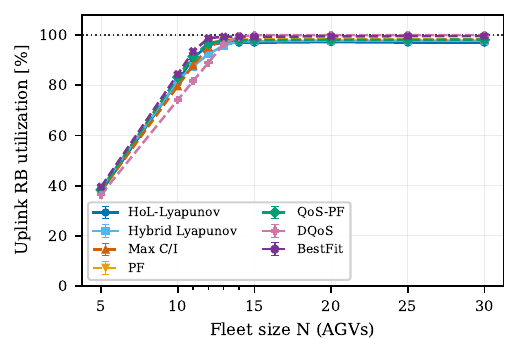}
\caption{Uplink resource-block utilization vs.\ fleet size $N$ (51 PRBs, 20 MHz). Saturation marks the capacity threshold behind the compliance knee.}
\label{fig:rb_util}
\end{figure}

With the fixed 51-PRB (20\,MHz) cell, the scenario supports the full traffic mix of approximately 12 AGVs. Through $N = 12$, every scheduler delivers $\ge 99.9\%$ of packets in all three classes (Fig.~\ref{fig:pdr}). Deadline compliance then collapses within the span of two additional vehicles: URLLC satisfaction at the relaxed 30\,ms target falls from 94.5\% to 12.3\% for DQoS and from 79.7\% to 0.7\% for HoL-Lyapunov between $N = 12$ and $N = 14$ (Fig.~\ref{fig:urllc_targets}), a sharp capacity \emph{threshold} rather than a gradual knee. Uplink resource-block utilization (Fig.~\ref{fig:rb_util}) confirms that the threshold is a genuine capacity limit: utilization rises from 36--40\% at $N = 5$ through 89--99\% at $N = 12$ and saturates at 97--100\% for every scheduler from $N = 14$ onward. All schedulers therefore face an identical resource budget, and the differences reported below reflect allocation policy alone.

\subsection{URLLC Latency Floor and Relaxed Targets}
\label{sec:latency_floor}
The strict 10\,ms URLLC budget proved to lie below the latency floor of the simulated end-to-end system. Even at the lightest load ($N = 5$), all seven schedulers exhibit URLLC mean latencies of $\approx 9$--11\,ms and 99th percentiles of $\approx 30$--38\,ms, and none exceeds 81\% compliance at 10\,ms. The floor does not arise from contention for radio resources, since fewer than 40\% of the resource blocks are scheduled at this load (Fig.~\ref{fig:rb_util}), nor from URLLC packets waiting behind eMBB data, since each class is held in its own DRB queue and the UE serves the URLLC bearer first when it fills a grant. The remaining contributions lie in the uplink access path that every policy shares: a URLLC packet waits until its UE has reported the backlog and received a grant, and a transmission that coincides with a clutter blockage may require retransmission. These contributions were not separated in this study. The gNB policy shortens the floor but does not remove it, as the head-of-line term lowers the 10\,ms deadline-miss ratio at $N = 5$ from 35.8\% to 22.0\% (Section~\ref{sec:ablation}) while no scheduler meets the budget. URLLC compliance is therefore additionally evaluated at post-hoc end-to-end targets of 15, 20, and 30\,ms (Fig.~\ref{fig:urllc_targets}); the scheduler ranking is consistent across targets.

\subsection{Feasible-Load Regime}
\label{sec:feasible}
Table~\ref{tab:supported} quantifies each scheduler's supported fleet size as the largest tested $N$ meeting a given compliance criterion, and Table~\ref{tab:nominal} reports the full per-class metric set at the nominal operating point $N = 12$. Below the threshold, the delay-aware baselines sustain the largest compliant fleets: DQoS meets the 30\,ms URLLC target at $\ge 95\%$ satisfaction up to $N = 11$ (and at $\ge 90\%$ up to $N = 12$), Max~C/I up to $N = 11$, while HoL-Lyapunov, PF, and QoS-PF support $N = 10$. At $N = 12$, DQoS also attains the lowest URLLC mean latency ($12.9 \pm 0.1$\,ms) and jitter ($11.9 \pm 0.6$\,ms). The margin between the best baseline and the proposed HoL-Lyapunov in this regime is thus a single vehicle, a deliberate trade-off in which the drift-plus-penalty structure optimizes queue stability rather than minimum delay, and the return on this concession appears past the threshold (Section~\ref{sec:overload}).

Fig.~\ref{fig:qos_sat} shows the same ordering in QoS satisfaction. At $N \le 10$, PF and QoS-PF attain the highest URLLC satisfaction, with HoL-Lyapunov within 4 percentage points of them; at $N = 13$ and $14$, DQoS and Max~C/I attain the highest satisfaction in every class. Past the threshold, the QoS satisfaction of HoL-Lyapunov falls to zero because the traffic it preserves is delivered after its budget (Table~\ref{tab:overload}); its overload advantage lies in delivery ratio and tail latency (Section~\ref{sec:overload}), not in deadline compliance.

\begin{table*}[t]
\centering
\caption{Supported fleet size: largest tested $N$ meeting each URLLC compliance criterion at the given end-to-end latency target (resolution limited to the tested fleet sizes $N \in \{5, 10, 11, \dots, 15, 20, 25, 30\}$). (a) URLLC evaluated at the 30 ms E2E target, eMBB/mMTC at their 100/300 ms budgets.}
\label{tab:supported}
\begin{tabular}{lrrrrrrr}
\toprule
Scheduler & \multicolumn{4}{c}{Max $N$, URLLC QoS-sat $\geq 95\%$} & \multicolumn{2}{c}{$\geq 90\%$} & Max $N$, all \\
 & 10 ms & 15 ms & 20 ms & 30 ms & 20 ms & 30 ms & classes $\geq 95\%$ \\
\midrule
HoL-Lyapunov & $<$5 & $<$5 & 5 & 10 & 10 & 11 & 10 \\
Hybrid Lyapunov & $<$5 & $<$5 & 5 & 5 & 5 & 10 & 5 \\
Max C/I & $<$5 & $<$5 & 5 & 11 & 11 & 11 & 11 \\
PF & $<$5 & $<$5 & 5 & 10 & 10 & 11 & 10 \\
QoS-PF & $<$5 & $<$5 & 5 & 10 & 10 & 11 & 10 \\
DQoS & $<$5 & $<$5 & 5 & 11 & 11 & 12 & 11 \\
BestFit & $<$5 & $<$5 & $<$5 & 5 & $<$5 & 10 & 5 \\
\bottomrule
\end{tabular}
\end{table*}

\subsection{Impact of the Head-of-Line Term}
\label{sec:ablation}
\begin{figure*}[t]
\centering
\includegraphics[width=\textwidth]{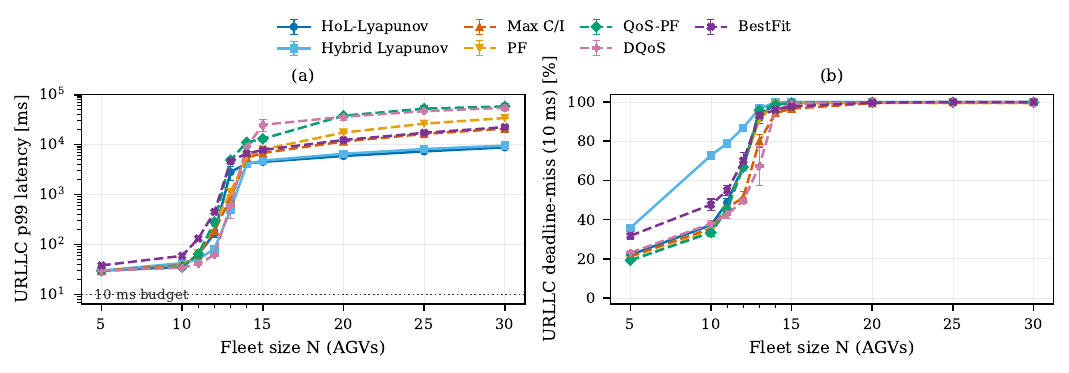}
\caption{URLLC tail behavior vs.\ fleet size $N$: (a) 99th-percentile latency (log scale, dotted line: 10 ms budget); (b) deadline-miss ratio at the 10 ms budget among delivered packets.}
\label{fig:urllc_tail}
\end{figure*}

The comparison between the two proposed variants isolates the contribution of the head-of-line delay term (Fig.~\ref{fig:urllc_tail}b). At $N = 10$, the plain Hybrid Lyapunov scheduler misses the 10\,ms deadline on $72.8 \pm 0.4\%$ of delivered URLLC packets, versus $37.3 \pm 1.2\%$ for HoL-Lyapunov, a 49\% reduction obtained purely by prioritizing head-of-line delay within the same drift framework (35.8\% vs.\ 22.0\% at $N = 5$). The plain variant's queue-length-driven metric is agnostic to how long the oldest packet has waited, which the URLLC tail exposes directly: its 10\,ms-target satisfaction at $N = 10$ is 27.2\%, versus 62.7\% for the HoL-enhanced variant. The HoL term is therefore essential, not incremental, for deadline-constrained traffic.

\subsection{Overload Regime}
\label{sec:overload}
Past the threshold, where no scheduler can meet URLLC deadlines (Section~\ref{sec:capacity_threshold}), the schedulers differ sharply in what they preserve (Fig.~\ref{fig:pdr}, Table~\ref{tab:overload}). The Lyapunov schedulers keep the prioritized classes' queueing systems preserved: at $N = 30$ ($2.5\times$ the supported fleet), HoL-Lyapunov still delivers $82.7 \pm 0.1\%$ of URLLC and $82.5 \pm 0.3\%$ of mMTC traffic with a bounded 99th-percentile latency of $8.8 \pm 0.04$\,s. The channel and allocation-oriented baselines form an intermediate group that partially protects the critical classes: Max~C/I delivers $65.7 \pm 0.1\%$ of URLLC (p99 $21.0$\,s) and BestFit $64.2 \pm 0.2\%$ (p99 $22.5$\,s), both at the expense of eMBB (20.0\% and 19.0\% PDR respectively). The PF / QoS-PF / DQoS family, by contrast, falls to 44--46\% with p99 tails of 34--58\,s and degrades all three classes roughly uniformly (e.g., DQoS at $N = 30$: 44--46\% PDR in every class), i.e., it distributes the shortfall across the traffic mix rather than resolving it in favour of the critical classes.

The counterpart is the collapse of the eMBB class (HoL-Lyapunov: 0\% eMBB PDR at $N \ge 25$; plain Lyapunov: $3.8\%$ at $N = 25$ and $0.4\%$ at $N = 30$). Two mechanisms, one intended and one incidental, produce it. The intended one is the scheduling order: the URLLC firewall term places the 10\,ms class ahead of both others, so under a $2.5\times$ demand excess the remaining classes contend for a residue. Note that this ordering does not rank mMTC above eMBB  the per-class weight of eMBB (priority~2, GBR) exceeds that of mMTC (priority~3, non-GBR) by a factor of four, and the head-of-line term grows three times faster for eMBB's 100\,ms budget than for mMTC's 300\,ms budget. The asymmetry in the results is one of demand, not of precedence: at $N = 30$ the offered load is 24\,Mb/s (URLLC), 118.3\,Mb/s (eMBB), and 0.50\,Mb/s (mMTC), so mMTC represents 0.35\% of the total and is satisfiable from the residue that URLLC leaves, whereas eMBB's demand exceeds the entire cell capacity and cannot be met at any position in the ordering. The incidental mechanism explains why eMBB delivery reaches \emph{exactly} zero rather than degrading proportionally: eMBB datagrams (12.4--60\,kB) are IP-fragmented into 9--41 fragments and count as delivered only if every fragment arrives before the reassembly timeout. A class served only from a residue transmits fragments too sparsely for reassembly to complete, so its delivery ratio falls discontinuously to zero once the residue drops below the level needed to clear whole datagrams.

This second mechanism carries an aggregate cost that must be stated plainly. All schedulers saturate the uplink at $N = 30$ (97--100\% RB utilization), yet total delivered goodput differs by a factor of three: 20.2\,Mb/s for HoL-Lyapunov and 18.5\,Mb/s for plain Lyapunov, against 35.6\,Mb/s for BestFit, 37.0\,Mb/s for Max~C/I, and 63.7\,Mb/s for PF and DQoS. The Lyapunov schedulers do not convert the lost eMBB throughput into equivalent gains for the critical classes: URLLC delivery improves by only $\approx 9$\,Mb/s while $\approx 52$\,Mb/s of eMBB goodput is lost. This gap arises because part of the spectrum is consumed by eMBB fragments that are successfully transmitted at the link level but later discarded during packet reassembly; thus, partial service of fragmented elastic flows consumes radio resources without contributing to delivered throughput. Therefore, the trade-off introduced by the proposed scheduler under overload is not simply ``critical traffic in exchange for elastic traffic.'' Instead, it prioritizes critical traffic reliability, nearly doubling URLLC delivery compared with the PF family and reducing the critical-tail performance degradation, but at the cost of approximately a two-thirds reduction in aggregate cell throughput. For a safety-critical industrial cell this is a defensible exchange: the sacrificed capacity would have served a class that, at $2.5\times$ overload, cannot be delivered usefully in any case, but it is a consequence of the interaction between residual service and IP reassembly, not a designed property of the drift formulation. Eliminating the waste requires the scheduler to withhold service from the elastic class outright rather than to rank it last, which we identify as future work (Section~\ref{sec:conclusion&futureWork}).

Latency statistics under overload must be read jointly with PDR (Table~\ref{tab:overload}). At $N = 30$, DQoS reports a lower mean URLLC latency than HoL-Lyapunov ($2.0$\,s vs.\ $7.6$\,s), yet it delivers barely half as many packets (45.6\% vs.\ 82.7\%) with a $6\times$ longer 99th percentile ($54.6$\,s vs.\ $8.8$\,s): its queues shed exactly the packets that would have raised the mean. This survivorship effect is why QoS satisfaction, which is computed over generated packets, is used as the headline metric throughout (Fig.~\ref{fig:qos_sat}).

\begin{figure*}[t]
\centering
\includegraphics[width=\textwidth]{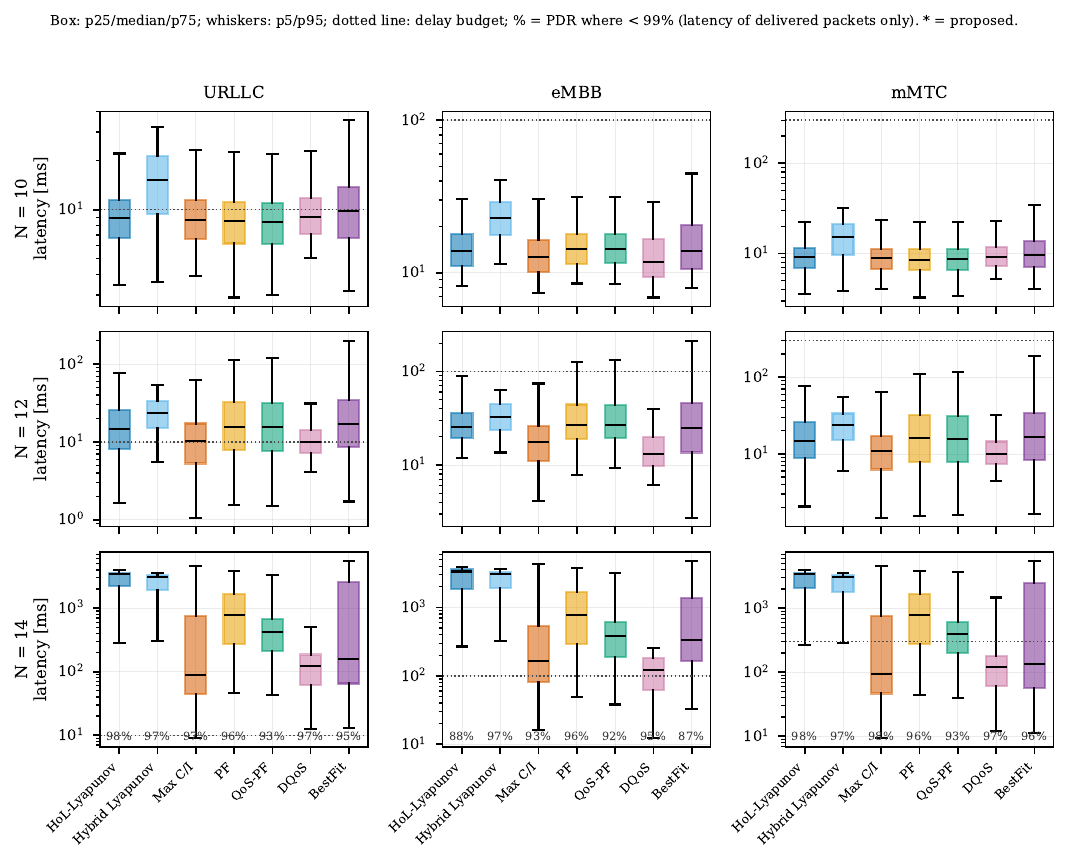}
\caption{End-to-end latency distribution of delivered packets at $N = 10, 12, 14$ AGVs (box: 25th/50th/75th percentiles; whiskers: 5th/95th; dotted line: class delay budget). Percentages below boxes give the PDR where it falls under 99\%: latency statistics cover delivered packets only, so a low-PDR scheduler can show deceptively low latency.}
\label{fig:latency_box}
\end{figure*}

\subsection{Summary of Findings}
\label{sec:findings}
The campaign supports three quantitative claims. First, on a fixed 20\,MHz industrial allocation the scenario's capacity threshold sits at $N \approx 12$ AGVs, confirmed by resource saturation, and is common to all scheduling policies. Second, within the feasible regime the proposed HoL-Lyapunov scheduler is competitive with the strongest baselines (supported fleet of 10 vs.\ 11 AGVs at the 30\,ms/95\% criterion), and its head-of-line term halves the URLLC deadline-miss ratio relative to the plain Lyapunov formulation. Third, under overload the proposed scheduler delivers $1.8\times$ the URLLC traffic compared to the PF / QoS-PF / DQoS family with a $4$--$7\times$ shorter 99th-percentile tail, resolving the capacity shortfall in favour of the critical classes rather than spreading it across the traffic mix, the property of primary operational value whenever offered load exceeds the available capacity, whether because the active fleet has grown or because the channel has degraded. This robustness is obtained at a measured cost in aggregate cell throughput (Section~\ref{sec:overload}).

\section{Conclusion and Future Work}
\label{sec:conclusion&futureWork}

This paper presented the HoL-Enhanced Hybrid Lyapunov scheduler, a 5G NR MAC scheduling policy that combines drift-plus-penalty queue stability with explicit head-of-line delay awareness and class isolation, and evaluated it against various stock Simu5G and SOTA schedulers in a physically detailed 3GPP Indoor Factory scenario with per-class QoS-flow bearers. The fleet-size sweep yielded three findings. First, the scenario exhibits a scheduler-independent capacity threshold at approximately 12 AGVs on the fixed 20\,MHz allocation, confirmed by uplink resource-block utilization saturating from 36--40\% at 5 vehicles to 97--100\% at 14: no scheduling policy can move the threshold, only shape behavior on either side of it. Second, within the feasible regime the proposed scheduler is competitive with the strongest delay-aware baselines supporting 10 AGVs against 11 for DQoS and Max~C/I at the 30\,ms/95\% URLLC criterion and the ablation against the plain Hybrid Lyapunov variant showed that the head-of-line term is essential rather than incremental, halving the URLLC deadline-miss ratio at nominal load. Third, and most consequentially, under overload the proposed scheduler preserves the traffic that matters: at $2.5\times$ the supported fleet it still delivers 82.7\% of URLLC and 82.5\% of mMTC packets with a bounded 99th-percentile latency, while the proportional-fair and delay-budget-aware baselines fall to 44--46\% delivery in every class with tails of tens of seconds. This graceful degradation resolves the shortfall in favour of the critical classes rather than spreading it uniformly; its counterpart is the collapse of the elastic eMBB class and, because partially served fragmented datagrams fail reassembly, a two-thirds reduction in aggregate delivered throughput  a defensible exchange for a safety-critical cell, and one that a service-withholding admission rule could largely recover.

These results carry three implications for industrial 5G and, by extension, 6G deployments. First, capacity planning and scheduler selection are separable problems: the admission limit of a private industrial cell is set by spectrum and traffic volume, and operators should dimension the fleet against the measured threshold rather than expect a scheduler to extend it. The scheduler's role begins where planning ends in the transient or fault-induced episodes when offered load exceeds the plan and there the choice of policy changes outcomes by a factor of two in delivered critical traffic. A practical deployment could therefore pair a delay-aware policy in the nominal regime with a Lyapunov-based policy under detected congestion, or adopt the proposed scheduler throughout as insurance at the cost of roughly one vehicle of nominal headroom. Second, the latency floor that no policy overcame even at minimal load shows that, in the evaluated configuration, sub-10\,ms end-to-end guarantees cannot be delivered by dynamic uplink scheduling alone. Mechanisms that shorten the uplink access path of critical traffic, such as configured grants or dedicated mini-slot resources, are natural complements, although the contribution of blockage-induced retransmission to the floor remains to be quantified. Third, overload evaluation must use survivorship-immune metrics: mean latency conditioned on delivery rewarded the policies that dropped the most packets, and QoS satisfaction computed over generated traffic reversed that ranking.

Several directions follow from this work. The most immediate is the adaptive tuning of the drift-plus-penalty weights $\alpha$, $\beta$, and $\gamma$, which were held fixed across the entire load range in this study: a lightweight learning agent e.g., contextual bandits or deep reinforcement learning driven by observed queue backlogs, head-of-line delays, and resource utilization could adjust the weights online, sharpening delay performance in the feasible regime while preserving the stability-oriented behavior under overload, and potentially recovering the one-vehicle gap to the delay-aware baselines without forfeiting overload robustness. A second and more immediate one follows from the throughput accounting of Section~\ref{sec:overload}: because a fragmented elastic flow served from a residue consumes spectrum on fragments that never complete reassembly, ranking the elastic class last is measurably worse than withholding service from it outright. Coupling the scheduler to an explicit congestion-triggered admission rule suspending the elastic class once the critical backlog exceeds a drift threshold, rather than merely deprioritizing it should recover a large part of the observed two-thirds throughput loss without altering the critical-class guarantees, and quantifying that gain is a direct extension of this work. A third direction is to decompose the latency floor identified above into its uplink components, separating the wait for a grant from retransmissions caused by blockage, and to reduce the dominant one, for example by combining the scheduler with configured grants for the critical DRB. Fourth, the evaluation should be broadened toward deployment realism: multi-cell layouts with inter-cell interference and handover, larger bandwidths and FR2 carriers, and end-to-end integration with TSN schedules so that radio-side guarantees compose with IEEE~802.1Qbv gating across the bridged domain. Finally, validating the two-regime behavior on a hardware testbed with physical AGVs would confirm that the overload-robustness advantage demonstrated here in simulation carries over to real industrial radio conditions.

\section{Acknowledgments}
This publication has emanated from research conducted with the financial support of Taighde Eireann Research Ireland under Grant number 13/RC/2077 P2 (CONNECT: the Research Ireland Centre for Future Networks). For the purpose of Open Access, the author has applied a CC-BY public copyright licence to any Author Accepted Manuscript version arising from this submission.

\bibliography{references}
\bibliographystyle{IEEEtran}

\end{document}